\documentclass[twocolumn,showpacs,superscriptaddress,floatfix,prb]{revtex4}

\usepackage{color} 
\usepackage{tabularx} 
\usepackage{epsfig}
\usepackage{amsmath} 
\usepackage{amssymb} 
\usepackage{graphicx}
\usepackage{enumerate}

\begin{document}

\title{Universality and universal finite-size scaling functions\\
  in four-dimensional Ising spin glasses}

\author{Thomas J\"org} \affiliation{Laboratoire de Physique Th\'eorique et
  Mod\`eles Statistiques, Universit\'e de Paris-Sud, b\^atiment 100, 91405
  Orsay Cedex, France} \affiliation{\'Equipe TAO - INRIA Futurs, 91405 Orsay
  Cedex, France}

\author{Helmut G.~Katzgraber} \affiliation {Theoretische Physik, ETH
  Zurich, CH-8093 Zurich, Switzerland}

\date{\today}

\begin{abstract}

We study the four-dimensional Ising spin glass with Gaussian and
bond-diluted bimodal distributed interactions via large-scale Monte Carlo
simulations and show via an extensive finite-size scaling analysis that
four-dimensional Ising spin glasses obey universality.

\end{abstract}

\pacs{75.50.Lk, 75.40.Mg, 05.50.+q}
\maketitle

\section{Introduction}
\label{sec:introduction}

The concept of universality, according to which the values of different
quantities, such as for example critical exponents, do not depend on
the microscopic details of a model, is well established in the theory
of critical phenomena of systems without frustration. However, for spin
glasses,\cite{binder:86} which have both frustration and disorder, the
situation has been less clear until recently.  Large-scale simulations
in three space dimensions\cite{joerg:06,katzgraber:06,hasenbusch:08}
for different disorder distributions of the random interactions
between the spins have shown that the critical exponents, the values
of different observables at criticality, as well as the finite-size
scaling functions---all which are necessary ingredients to show
that different models are in the same universality class---agree
well within error bars.  The conclusion that universality holds
has also been obtained via other methods such as high-temperature
series expansions\cite{daboul:04} where the critical exponent
$\gamma$ has been studied. Studies of dynamical quantities however
have yielded different critical exponents for different disorder
distributions,\cite{mari:99,campbell:00,mari:01,pleimling:05} although
it is unclear up to what level it can be expected that ``dynamical
universality'' can be compared to universality in thermal equilibrium.

In two space dimensions things are less clear: Because the transition
to a spin-glass phase only happens at zero temperature, it is believed
that systems with discrete and continuous coupling distributions
behave in a different way.\cite{amoruso:03} In particular, the
ground-state entropy is nonzero in the former while it is zero in the
latter. Thus one might expect that at the zero-temperature critical
point the critical exponent $\eta$ is different.  Recently, however,
an alternate scenario for universality in two space dimensions has
been proposed,\cite{joerg:06a} where it is expected that as long as
the temperature is nonzero two-dimensional spin glasses with different
disorder distributions belong to the same universality class and where
the different behavior seen at zero temperature is explained by the
presence of an additional spurious fixed-point.\cite{comment:dilute_FP}
Limitations in the simulation techniques and analysis methods
have so far yielded no conclusive results making this proposal
controversial.\cite{katzgraber:07c,hartmann:08}

In spin glasses it is extremely difficult and numerically very costly
to determine critical exponents with high precision. This is mainly
due to the following reason: It is difficult to sample the disorder
average with good enough statistics, especially for large system
sizes, because spin glasses have very long equilibration times
in Monte Carlo simulations and as a consequence one has to deal
with corrections to scaling due to a very limited range of system
sizes at hand.  In previous studies\cite{katzgraber:06,joerg:06}
only statistical error bars had been considered. Because of limited
system sizes and corrections to scaling in three space dimensions,
deviations between the critical parameters beyond statistical error
bars can be expected and indeed this expectation has very recently
been confirmed in Ref.~\onlinecite{hasenbusch:08} in a very thorough
study where for the first time corrections to scaling have been
studied with good accuracy. While studying higher-dimensional systems
than three space dimensions might seem paradoxical at first because
of the aforementioned problems, the proximity to the upper critical
dimension $d_{\rm ucd} = 6$ is advantageous.  High-temperature series
expansion studies\cite{daboul:04} suggest that corrections to scaling
should be falling off fast in four-dimensional Ising spin glasses,
i.e., that the leading correction-to-scaling exponent $\omega$ is
large. More specifically, in Ref.~\onlinecite{daboul:04} a value
for $\omega$ between $1.3$ and $1.6$ was found. Although the range
of system sizes accessible to Monte Carlo simulations in four space
dimensions is more limited than in three space dimensions, the model
poses a ``good compromise'' case where corrections can be kept small
while the system sizes are reasonably large.  In this article we
choose the observables that display the smallest corrections to
scaling for the system sizes accessible via Monte Carlo simulations.
We feel that introducing correction-to-scaling terms along the lines
of Ref.~\onlinecite{beach:05} in order to give more precise estimates
for the critical exponents cannot be controlled sufficiently well,
and thus, we do not display an analysis using these methods. However,
we show that our results are perfectly compatible with a large
correction-to-scaling exponent $\omega$.

As with the three-dimensional Ising spin
glass,\cite{kawashima:03,kawashima:04,katzgraber:06} there have been
many different estimates for the critical exponents (especially for
the anomalous dimension $\eta$) of the four-dimensional Ising spin
glass as shown in Table \ref{tab:previous}.

\begin{table*}
\caption{
Different estimates (sorted chronologically and with respect to the
method used for the determination) of the critical exponents computed
by different groups for Gaussian (G), bimodal ($\pm J$), triangular
(T), uniform (U) and Laplacian (L) random bonds. The estimates,
especially the ones for $\eta$, show strong variations and often
do not agree. The last two rows show the results from this study.
The critical temperatures denoted by an asterisk indicate that the
variance of the coupling distribution used in the corresponding study
is not normalized to unity (Ref.~\onlinecite{comment:wang}).
\label{tab:previous}}
\begin{tabular*}{\textwidth}{@{\extracolsep{\fill}} l l l l l l l }
  \hline
  \hline
  Authors                                                     & Couplings & Method & $\gamma$ & $\nu$ & $\eta$ & $T_c$ \\
  \hline
  Singh \& Chakravarty (Ref.~\onlinecite{singh:86})           & $\pm J$   & Series & $2.0(4)$ & & & $2.02(6)$ \\
  Klein {\em et al.} (Ref.~\onlinecite{klein:91})             & $\pm J$   &        & $2.00(25)$ & $0.95$ & $-0.11$ & $2.04(3)$ \\
  Daboul {\em et al.} (Ref.~\onlinecite{daboul:04})           & U         &        & $2.4(1)$ & & & $1.10(2)^*$ \\
                                                              & $\pm J$   &        & $2.5(3)$   &    &    &  $1.96(7)$   \\
                                                              & G         &        & $2.3(1)$   &    &    &  $1.79(1)$   \\
                                                              & T         &        & $2.5(2)$   &    &    &  $1.36(3)^*$ \\
  \hline
  Bernardi \& Campbell (Ref.~\onlinecite{bernardi:97})        & $\pm J$   & Dynamic MC & & & $-0.31(1)$ & $1.99(1)$ \\
                                                              & G         &            &   &  & $-0.47(2)$ & $1.77(1)$ \\
                                                              & U         &            &   &  & $-0.37(2)$ & $1.91(1)$ \\
                                                              & L         &            &   &  & $-0.60(3)$ & $1.52(1)$ \\
  \hline
  Bhatt \& Young (Ref.~\onlinecite{bhatt:88})                 & G         & Static MC & $1.8(4)$ & $0.8$ & $-0.30(15)$ & $1.75(5)$ \\
  Reger {\em et al.} (Ref.~\onlinecite{reger:90})             & $\pm J$   & & & & $-0.5$ & \\
  Parisi {\em et al.} (Ref.~\onlinecite{parisi:96})           & G         & & $2.1(2)$ & $0.9(1)$ & $-0.35(5)$ & $1.80(1)$ \\
  Ney-Nifle (Ref.~\onlinecite{neynifle:98})                   & G         & & & $0.87(15)$ & & $1.80(5)$ \\
  Marinari \& Zuliani (Ref.~\onlinecite{marinari:99})         & $\pm J$   & & & $1.0(1)$ & $-0.30(5)$ & $2.03(3)$ \\
  Hukushima (Ref.~\onlinecite{hukushima:99})                  & $\pm J$   & & & $0.92(6)$ & & $2.00(4)$ \\
  \hline                                                      
  This study                                                  & G & Static MC & $2.32(8)$ & $1.02(2)$ & $-0.275(25)$ & $1.805(10)$ \\
                                                              & $\pm J$ bond-diluted & & $2.33(6)$ & $1.025(15)$ & $-0.275(25)$ & $1.0385(25)^*$ \\
  \hline
  \hline
\end{tabular*}
\end{table*}

The main conclusion of this work is that equilibrium universality in
four-dimensional spin glasses is satisfied, since we find agreement
within error bars for all the finite-size scaling functions and
critical exponents studied.

In Sec.~\ref{sec:model} we introduce the model as well as the measured
observables. In addition, we describe the numerical methods used.
Results are presented in Sec.~\ref{sec:results}, followed by concluding
remarks in Sec.~\ref{sec:conclusions}. Finally, a discussion of
other commonly-used observables that have been less useful in the
present analysis is given in Appendix \ref{sec:appendix}. Details
of the analysis are presented in Appendices \ref{app:derivation}
and \ref{app:quotient}, as well as an extended scaling analysis of
the data with Gaussian disorder in Appendix \ref{app:campbell}.

\section{Model, observables, and numerical details}
\label{sec:model}

\subsection{Edwards-Anderson model}
\label{subsec:model}

The Edwards-Anderson (EA) Ising spin-glass
Hamiltonian\cite{edwards:75,binder:86} is given by
\begin{equation} {\mathcal H} = -\sum_{\langle i,j \rangle} J_{ij} S_i S_j .
  \label{eq:ham}
\end{equation}
The Ising spins $S_i = \pm1$ lie on a hypercubic lattice of size $N =
L^d$ in $d = 4$ space dimensions with periodic boundary conditions. The
sum is over nearest neighbors on the lattice. We study two versions
of the model:
\begin{enumerate}[(i)]
\item{Gaussian-distributed interactions $J_{ij}$ with zero mean and standard
deviation unity;
    \begin{equation}
      {\mathcal P}(J_{ij}) = \frac{1}{\sqrt{2\pi}}\, e^{-J_{ij}^2/2}.
      \label{eq:model_gauss}
    \end{equation}
  }
\item{Bimodal-distributed random interactions\cite{shapira:94} with a 
bond dilution of $65\%$, i.e.,
    \begin{equation}
      \begin{split}
        {\mathcal P}(J_{ij}) &= \,(1-p) \, \delta (J_{ij}) \\ &\quad+
        \frac{p}{2} \, \left[ \delta (J_{ij} - 1) + \delta(J_{ij} + 1)
        \right]\!,
        \label{eq:model_pmjdil}
      \end{split}
    \end{equation}
with $p=0.35$ (Ref.~\onlinecite{comment:dil}).}
\end{enumerate}

\subsection{Measured observables}
\label{subsec:observables}

In order to compute the critical parameters and henceforth test
for universality, we compute different observables that are
known to show a good signature of the phase transition. The Binder
cumulant,\cite{binder:81} defined via
\begin{equation}
  g(L,T) = \frac{1}{2}
  \left(
    3 - \frac{[{\langle q^4\rangle}]_{\rm av}}
	     {[{\langle q^2\rangle}]_{\rm av}^2}
  \right)\!,
  \label{eq:Binder}
\end{equation}
is dimensionless, and scales as
\begin{equation}
  g(L,T) = \widetilde{g}[B L^{1/\nu}(T - T_{c})] +  {\rm corrections}.
  \label{eq:g_scale}
\end{equation}
Here $T_{c}$ is the critical temperature and $B$ is a metric
factor.  The critical exponent $\nu$ describes the divergence
of the infinite-volume correlation length $\xi(T)$ as the
temperature approaches $T_c$, i.e., $\xi(T) \sim |T-T_c|^{-\nu}$. The
corrections to scaling in Eq.~(\ref{eq:g_scale}), as well
as in Eqs.~(\ref{eq:xi_scale}) and (\ref{eq:chisg_scale}) are
asymptotically dominated by the leading correction-to-scaling exponent
$\omega$ and vanish in the thermodynamic limit ($L\to\infty$). In
Eq.~(\ref{eq:Binder}) $\langle {\mathcal O} \rangle$ represents
a thermal average of an observable ${\mathcal O}$, $[{{\mathcal
O}}]_{\rm av}$ is a disorder average, and $q$ is the spin overlap;
\begin{equation}
  q = \frac{1}{N}\sum_{i = 1}^N S_i^a S_i^b \! .
  \label{eq:q}
\end{equation}
In Eq.~(\ref{eq:q}) ``$a$'' and ``$b$'' represent two replicas of the
system with the same disorder. In addition, we study the finite-size
correlation length,\cite{cooper:82,palassini:99b,ballesteros:00}
\begin{equation}
  \xi(L,T) = \frac{1}{2 \sin (|{\bf k}_\mathrm{min}|/2)}
  \left[\frac{\chi_{\rm SG}({\bf 0})}{\chi_{\rm SG}({\bf k}_\mathrm{min})} 
    - 1\right]^{1/2}\!,
  \label{eq:xiL}
\end{equation}
where ${\bf k}_\mathrm{min} = (2\pi/L, 0, 0, 0)$ is the smallest
nonzero wave vector and the wave-vector-dependent spin-glass
susceptibility is given by
\begin{equation}
  \chi_{\rm SG}({\bf k}) = \frac{1}{N} \sum_{i, j} 
  [{\langle S_i S_j \rangle^2 }]_{\rm av} \;
  e^{\imath{\bf k}\cdot({\bf R}_i - {\bf R}_j)} \! .
  \label{eq:chisg}
\end{equation}
The finite-size correlation length $\xi(L,T)$ divided by the system
size is also a dimensionless quantity, i.e.,
\begin{equation}
  \frac{\xi(L,T)}{L} = \widetilde{\xi}[B L^{1/\nu}(T - T_{c})]  + {\rm
    corrections}.
  \label{eq:xi_scale}
\end{equation}
For $L\rightarrow \infty$ data for $\xi(L,T)/L$ as well as for $g(L,T)$
intersect at $T \rightarrow T_c$. For finite systems the data cross
at an effective critical temperature $T_c^*(L)$ that converges
asymptotically to $T_c$ as\cite{binder:81}
\begin{equation}
  T_c^*(L) - T_c \propto L^{-\omega-1/\nu} \! .
  \label{eq:effective_Tc}
\end{equation}
In the following we denote the value of an observable ${\mathcal O}$
measured at this effective critical point by
\begin{equation}
 {\mathcal O}^* \:\dot{=}\: {\mathcal O}[T_c^*(L)].
\end{equation}
The definition of the finite-size correlation length $\xi(L,T)$
in Eq.~\eqref{eq:xiL} involves in general the same leading
corrections to scaling as $\chi_{\rm SG}$, which in turn is given by
$\omega$. Furthermore, this definition is not unique\cite{salas:00}
and different definitions of $\xi(L,T)/L$ show differences of the
order of $L^{-2}$. Such differences may seem irrelevant, but for the
small systems sizes that can be accessed in spin-glass simulations
correction terms might actually be visible in the data.  Note that in
Eqs.~(\ref{eq:g_scale}), (\ref{eq:xi_scale}), and (\ref{eq:chisg_scale}),
$T_{c}$ and the metric factor $B$ are nonuniversal, but, since $B$ is 
included explicitly, the scaling functions $\widetilde{g}(x)$ and 
$\widetilde{\xi}(x)$ are both universal.\cite{privman:84,katzgraber:06} 
Since for both disorder distributions studied we use the same boundary 
conditions and sample shapes, these scaling functions are expected to
be the same for different disorder distributions if the systems are 
in the same universality class. This is a necessary yet not sufficient
condition. In addition, the critical exponents, as well as the values
of the scaling functions at criticality ($T = T_c$) have to agree.

In addition to studying the Binder cumulant as well as the finite-size
correlation length (from which we obtain $T_{c}$ and the critical
exponent $\nu$), we need to study another observable to obtain a {\em
second} critical exponent to fully characterize the universality class
of the model.\cite{yeomans:92} Thus we also study the scaling behavior
of the spin-glass susceptibility $\chi_{\rm SG} = \chi_{\rm SG}({\bf k}
= 0)$ (also $\chi_{\rm SG} = N[{\langle q^2 \rangle}]_{\rm av}$). The
spin-glass susceptibility scales as
\begin{equation}
  \chi_{\rm SG}(L,T) = C L^{2 - \eta} \widetilde{\chi}
  [B L^{1/\nu}(T - T_{c})] + {\rm corrections},
  \label{eq:chisg_scale}
\end{equation}
where the anomalous dimension $\eta$ is the second critical
exponent needed to establish the universality class of the model. In
Eq.~(\ref{eq:chisg_scale}) $C$ represents a nonuniversal amplitude. We
also study another related quantity $\zeta(L,T)$ that has shown to be
useful in determining the critical exponent $\eta$, which is defined
as\cite{calabrese:03}
\begin{equation}
  \zeta(L,T) = \frac{\chi_{\rm SG}}{\xi^2}. 
  \label{eq:zeta}
\end{equation}
The advantage of studying $\zeta$ is mainly given by the fact that
the statistical correlations between $\chi_{\rm SG}$ and $\xi$ lead
to smaller errors on $\zeta$ than on $\chi_{\rm SG}$ and therefore
to a more precise determination of $\eta$.

Finally, we study other phenomenological couplings that
have been suggested to compute the spin-glass transition
temperature, which are the lack of self-averaging $A$ given
by\cite{wiseman:95,aharony:96,wiseman:98,wiseman:98a,marinari:99b}
\begin{equation}
  A(L,T)=\frac{ [{\langle
      	q^2 \rangle^2}]_{\rm av} - 
	[{ \langle q^2\rangle}]_{\rm av}^2}{[{\langle q^2 \rangle}]_{\rm
          av}^2 }, 
      \label{eq:A}
\end{equation}
and the Guerra parameter $G$ given by\cite{guerra:96,marinari:98e}
\begin{equation}
  G(L,T)=\frac{[{\langle q^2\rangle^2}]_{\rm av}-[{\langle
      	q^2\rangle}]_{\rm av}^2} {[{\langle q^4\rangle}]_{\rm av}-
	[{\langle q^2\rangle}]_{\rm av}^2}.
  \label{eq:Guerra}
\end{equation}
Both $A$ and $G$ are related to the Binder cumulant $g$ through the
relation $g=1-A/(2G)$.

The scaling expressions in Eqs.~(\ref{eq:g_scale}), (\ref{eq:xi_scale}),
and (\ref{eq:chisg_scale}) can be used to determine the critical
exponents, but in practice this strategy is not very promising because
especially in the case of the spin-glass susceptibility analytic
corrections to scaling are very easily confused with the leading
scaling behavior, which in turn leads to unreliable estimates
of the critical exponents.\cite{katzgraber:06,hasenbusch:08}
In order to determine reliable estimates for the critical
quantities, we use the quotient method, which avoids the
problems of the analytic corrections to scaling in an elegant
manner.\cite{ballesteros:97,ballesteros:00,comment:campbell2}
For any observable ${\mathcal{O}}(L,T)$ and the 
finite-size correlation length, finite-size scaling theory
predicts\cite{fisher:72,privman:90,caracciolo:95} that
\begin{equation}
  \frac{{\mathcal O}(L,T)}{{\mathcal O}(\infty,T)}  =
  f_{\mathcal O} \Bigl[ \xi(\infty,T)/L \Bigr] +
  O\Bigl(\xi^{-\omega},L^{-\omega}\Bigr),
  \label{eq:FSS_1}
\end{equation}
as well as
\begin{equation}
  \frac{{\mathcal O}(s L,T)}{{\mathcal O}(L,T)}   =
  F_{{\mathcal O}} \Bigl[ \xi(L,T)/L; s \Bigr] +
  O\Bigl(\xi^{-\omega},L^{-\omega}\Bigr),
  \label{eq:FSS_2}
\end{equation}
where $f_{{\mathcal O}}$ and $F_{{\mathcal O}}$ are
universal finite-size scaling functions and $s > 1$ is a scale
factor. The exponent $\omega$ is again the leading nonanalytic
correction-to-scaling exponent. Because in Eq.~(\ref{eq:FSS_2})
only pairs of {\em finite} system sizes $L$ and $s L$ appear, this
formulation is well adapted for use in numerical simulations. For
example, the knowledge of the universal scaling functions $F_{\chi}$
and $F_{{\xi}}$ [meaning ${\mathcal O} = \chi$ or ${\mathcal O} =
\xi$ in Eq.~(\ref{eq:FSS_2}), respectively] allows us to extract the
critical exponents $\eta$ and $\nu$ using the quotient method. For
the quotient method one defines an effective critical temperature
$T^*_{c}$ at which the correlation length measured in units of the
lattice size $L$ is equal for the pair of systems, i.e.,
\begin{equation}
  \xi(L,T^*_{c})/L = \xi(s L,T^*_{c})/(sL),
  \label{eq:equal_correlation_length}
\end{equation}
or alternatively,
\begin{equation}
  g(L,T^*_{c}) = g(s L,T^*_{c}).
  \label{eq:equal_binder}
\end{equation}
Note that we do not use a different notation for $T^*_{c}$
defined through Eqs.~(\ref{eq:equal_correlation_length}) and
(\ref{eq:equal_binder}) although the corresponding $T^*_{c}$ in
general is different, and only in the thermodynamic limit converges
to a unique value $T_c$ as indicated in Eq.~(\ref{eq:effective_Tc}).
The crossings are determined by fitting a cubic spline through the
data.  We have refrained from using reweighting techniques for the
determination of the crossings,\cite{ferrenberg:88,ballesteros:97}
because the remaining statistical errors of the sample average dominate
the errors even for the good statistics we have at hand. In addition
to Eq.~(\ref{eq:FSS_2}) we also study the following equivalent relation
\begin{equation}
  \frac{{\mathcal O}(s L,T)}{{\mathcal O}(L,T)} =
  F_{{\mathcal O}} \Bigl[ g(L,T); s \Bigr] +
  O\Bigl(\xi^{-\omega},L^{-\omega}\Bigr) 
  \label{eq:FSS_with_binder}
\end{equation}
which, however, will prove to be advantageous in the present
study. Another case in which this version of the finite-size scaling
relation is very useful is to study scaling properties {\em within}
the spin-glass phase.\cite{joerg:08b}

The critical exponent $x_{\mathcal O}$ associated with a given observable
which at criticality diverges as $(T - T_{c})^{-x_{\mathcal O}}$
can then be estimated via the quotient
\begin{equation}
  s^{x_{\mathcal O}/\nu} = 
  \frac{{\mathcal O}(s L,T^*_c)}{{\mathcal O}(L,T^*_c)} + O(L^{-\omega}),
  \label{eq:quotient_1}
\end{equation}
and thus from the finite-size scaling function $F_{{\mathcal O}}$.
In Eq.~(\ref{eq:quotient_1}) the critical exponent $\nu$ of the
correlation length is unknown, but it can be estimated for example
from the finite-size scaling function of the temperature derivative
of the correlation length $\xi(L,T)$, $F_{{\partial_{T} \xi}}$, via
\begin{equation}
  s^{1/\nu} = 1 + \frac{x^*}{s} \partial_x F_{{\xi}}(x,s) \big|_{x=x^*}
  + O(L^{-\omega}),  
  \label{eq:quotient_2}
\end{equation}
with $x = \xi(L,T)/L$ and $x^* = \xi(L,T_c^*)/L$. Alternatively,
$\nu$ can also be estimated from the temperature derivative of the
finite-size scaling function of the Binder cumulant $F_{{\partial_{T}
g}}$, i.e., 
\begin{equation}
  s^{1/\nu} = 1 + g^* \partial_g F_{g}(g,s) \big|_{g=g^*}
  + O(L^{-\omega}).
  \label{eq:quotient_3}
\end{equation}
Here $g^* = g(T_c^*,L)$. In our study we fit cubic splines to
the data of $F_{\xi}$ and $F_{g}$ to calculate the derivatives in
Eqs.~(\ref{eq:quotient_2}) and (\ref{eq:quotient_3}). A detailed
derivation of Eqs.~\eqref{eq:quotient_2} and \eqref{eq:quotient_3}
is given in Appendix \ref{app:derivation}.

The anomalous dimension $\eta$ can be determined from $F_{\chi}$. Using
the scaling relation
\begin{equation}
  \gamma = \nu (2 - \eta),
  \label{eq:gamma_scaling}
\end{equation}
we obtain
\begin{equation}
  s^{\gamma/\nu} = s^{2 - \eta} = F_{\chi}^* + O(L^{-\omega}).
  \label{eq:Fchi}
\end{equation}
Similarly, $\eta$ can be obtained from the finite-size scaling
function $F_{\zeta}$,
\begin{equation}
  s^{-\eta} = F_{\zeta}^* + O(L^{-\omega}).
  \label{eq:Fzeta}
\end{equation}
We determine the critical exponent $\gamma$ from
Eq.~(\ref{eq:gamma_scaling}) using the estimates of $\nu$ and
$\eta$.  Finally, we also study the Binder cumulant as a function of
$\xi(L,T)/L$ (Ref.~\onlinecite{joerg:06,katzgraber:06,hasenbusch:08}
and \onlinecite{kim:96}).  Nonuniversal metric factors cancel and
so $g(L,T) = \widehat{g}[\xi(L,T)/L]$ with $\widehat{g}$ a universal
function.\cite{katzgraber:06,joerg:06} Therefore data for different
disorder distributions should fall on the same universal curve if
the models share the same universality class.

\subsection{Simulation details}
\label{subsec:simulation}

The simulations are done using exchange (parallel tempering)
Monte Carlo\cite{hukushima:96,marinari:96} and the simulation
parameters are presented in Tables \ref{tab:simparams_gauss}
and \ref{tab:simparams_pmj} for the Gaussian and bond-diluted
bimodal disorder distributions, respectively. For the Gaussian
disorder we test equilibration using the method presented
in Ref.~\onlinecite{katzgraber:01}. For the bond-diluted bimodal
disorder we use a multispin coded approach to speed up the simulation
in addition to a cluster updating routine,\cite{joerg:05,joerg:06}
which can substantially speed up equilibration when the system is
diluted and complements the parallel tempering Monte Carlo updates.
Although we did no systematic study, we have the impression that
in four dimensions the additional cluster updates are somewhat
less effective than in lower space dimensions. Equilibration in the
bond-diluted case is tested by a logarithmic binning of the data. Once
the last three bins for all observables agree within error bars, we
declare the system to be in thermal equilibrium.

\begin{table}
\caption{
Parameters of the simulations for the Gaussian-distributed disorder.
$L$ denotes the system size. $N_{\rm sa}$ is the number of samples
and $N_{\rm sw}$ is the total number of Monte Carlo sweeps performed
on a single sample for each of the $2 N_T$ replicas.  $T_{\rm min}$
and $T_{\rm max}$ are the lowest and highest temperatures simulated,
and $N_T$ is the number of temperatures used in the parallel tempering
method.  \label{tab:simparams_gauss}
}
\begin{tabular*}{\columnwidth}{@{\extracolsep{\fill}} r r r r r r }
  \hline
  \hline
  $L$  &  $N_{\rm sa}$  & $N_{\rm sw}$ & $N_T$ & $T_{\rm min}$ & $T_{\rm max}$\\
  \hline
  $3$  & $20000$ & $131072$ & $29$ & $1.400$ & $3.061$\\
  $4$  & $20000$ & $131072$ & $29$ & $1.400$ & $3.061$\\
  $5$  & $20000$ & $131072$ & $29$ & $1.400$ & $3.061$\\
  $6$  & $20000$ & $131072$ & $29$ & $1.400$ & $3.061$\\
  $8$  &  $3500$ & $524288$ & $29$ & $1.400$ & $3.061$\\
  $10$ &  $2000$ & $524288$ & $29$ & $1.400$ & $3.061$\\
  \hline
  \hline
\end{tabular*}
\end{table}

\begin{table}
\caption{
Parameters of the simulations for the bond-diluted bimodal disorder
distribution. For details see Table ~\ref{tab:simparams_gauss}.
\label{tab:simparams_pmj}
}
\begin{tabular*}{\columnwidth}{@{\extracolsep{\fill}} r r r r r r }
  \hline
  \hline
  $L$  &  $N_{\rm sa}$  & $N_{\rm sw}$ & $N_T$ & $T_{\rm min}$ & $T_{\rm max}$\\
  \hline
  $3$  & $102400$ & $20000$  & $11$ & $0.950$ & $1.800$\\
  $4$  & $107680$ & $40000$  & $11$ & $0.950$ & $1.800$\\
  $5$  & $101699$ & $40000$  & $11$ & $0.950$ & $1.800$\\
  $6$  & $101664$ & $40000$  & $11$ & $0.950$ & $1.800$\\
  $8$  &  $41408$ & $100000$ & $21$ & $0.950$ & $1.800$\\
  $10$ &  $24160$ & $100000$ & $21$ & $0.950$ & $1.800$\\
  \hline
  \hline
\end{tabular*}
\end{table}

\section{Results}
\label{sec:results}

\begin{figure*}[htbf]
\includegraphics[width=0.95\columnwidth]{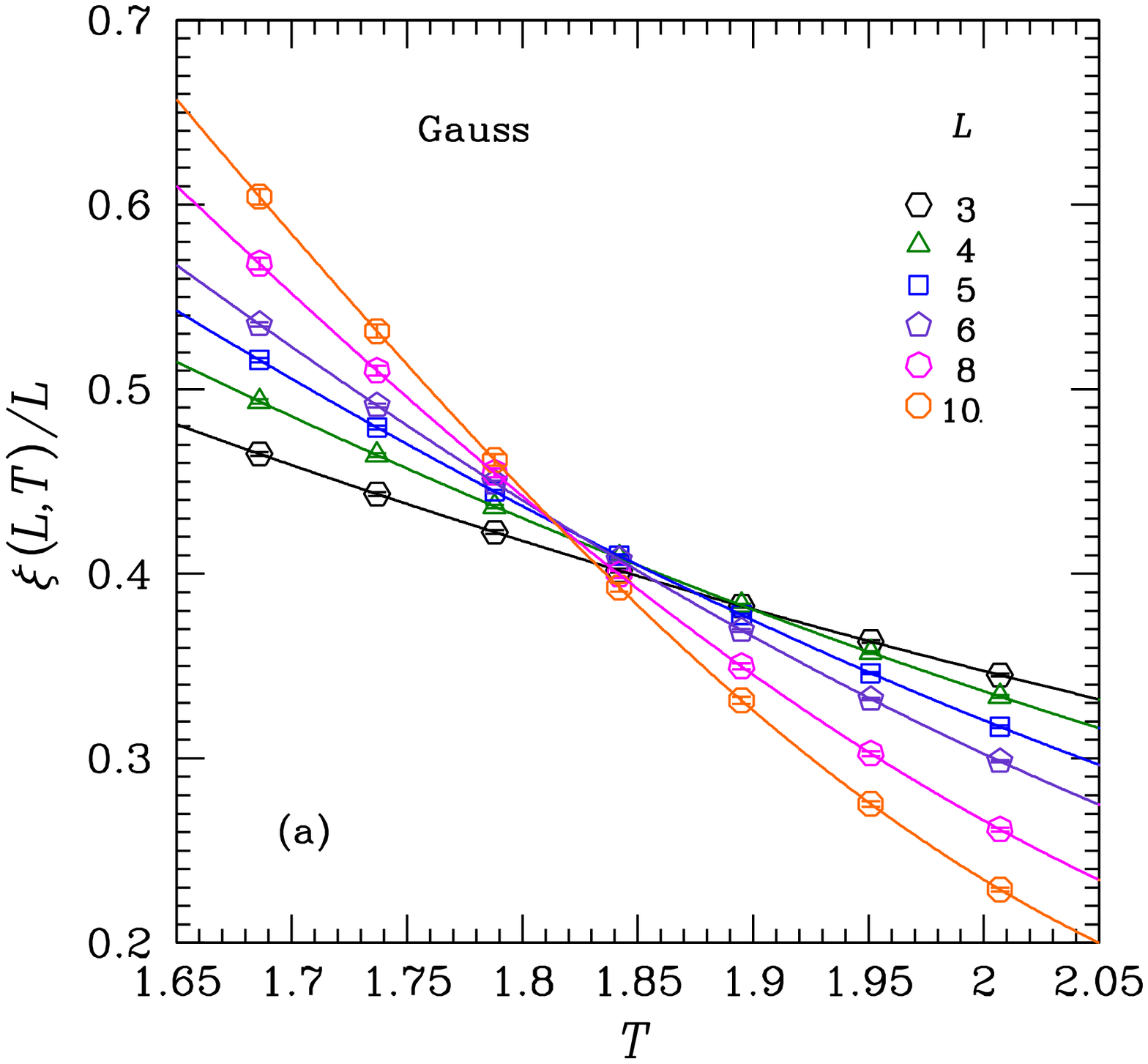}\quad
\includegraphics[width=0.95\columnwidth]{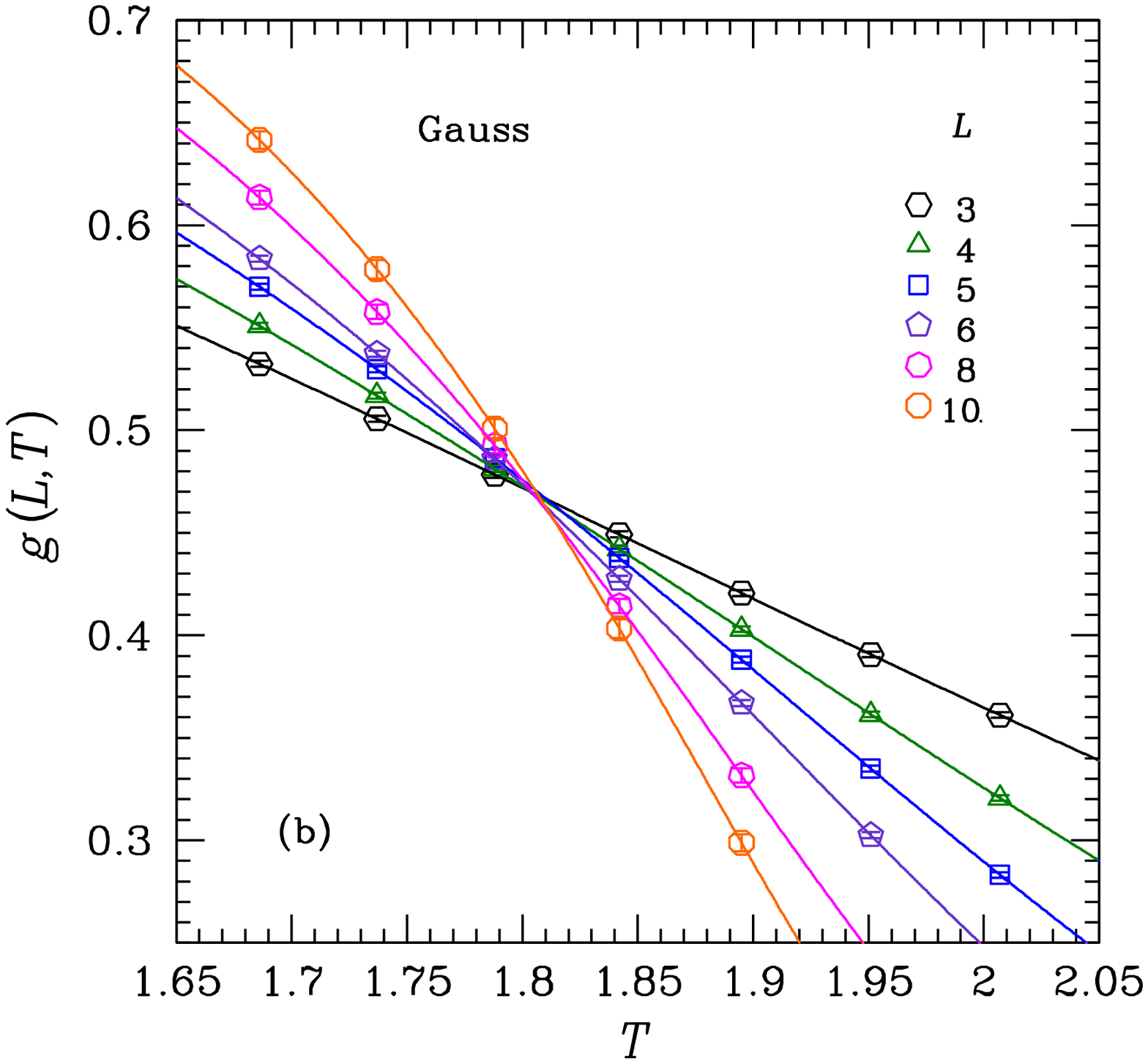}
\vspace*{-1.2cm}
\caption{(Color online) 
Determination of the critical temperature of the four-dimensional
Edwards-Anderson Ising spin glass with Gaussian disorder. In panel (a)
the finite-size correlation length $\xi(L,T)/L$ as a function of the
temperature $T$ for different system sizes $L$ is shown. The data for
$L\geq6$ cross at $T_c = 1.810(15)$.  In panel (b) the corresponding
data for the Binder cumulant $g(L,T)$ as a function of the temperature
$T$ for different system sizes $L$ are shown. The data cross at $T_c =
1.805(10)$, in agreement with the data for the correlation length.
The crossing of the data for the Binder cumulant is cleaner than for
the correlation length and shows no noticeable drift of the crossing
point with increasing system sizes.}
\label{fig:Tc_gauss}
\end{figure*}

In the following we perform a finite-size scaling analysis of different
observables for both models. We determine quantities with small scaling
corrections that provide good estimates for the critical temperature
and critical exponents. We then compare different finite-size scaling
functions for the model with different disorder distributions and
discuss the influence of corrections to scaling on our results.

\begin{figure*}
\includegraphics[width=0.95\columnwidth]{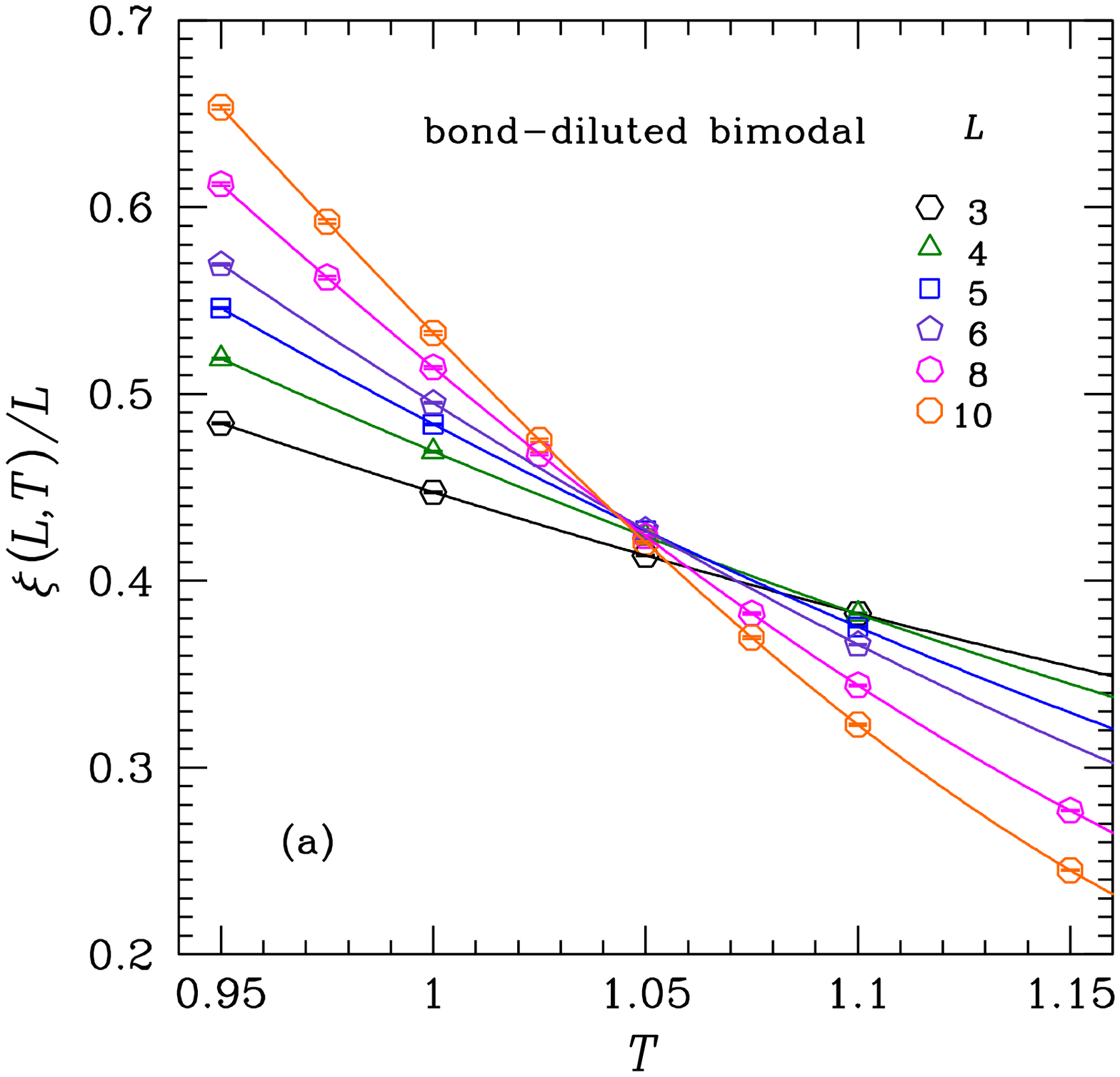}\quad
\includegraphics[width=0.95\columnwidth]{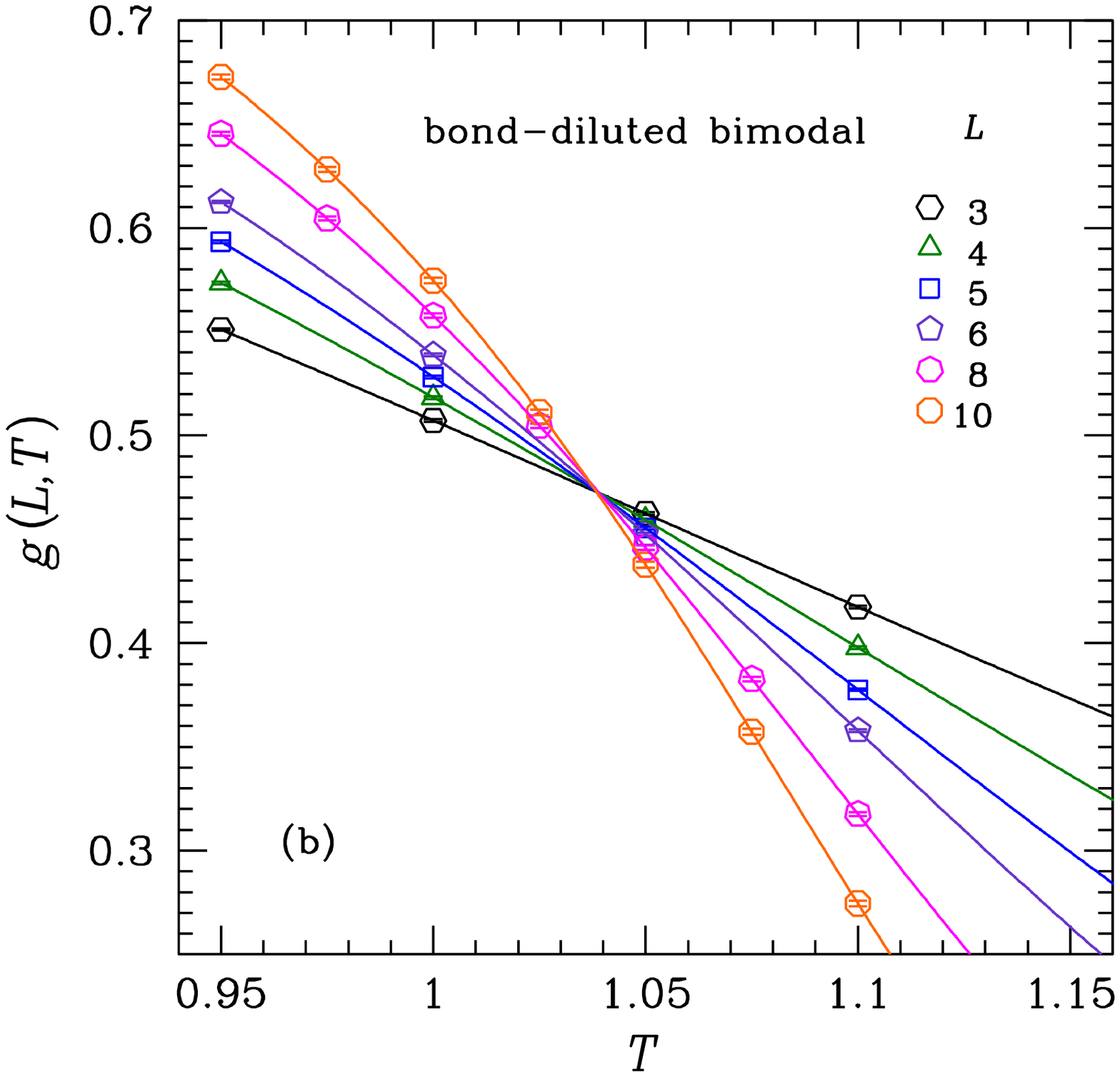}
\vspace*{-1.2cm}
\caption{(Color online) 
Determination of the critical temperature of the four-dimensional
Edwards-Anderson Ising spin glass with diluted bimodal disorder.
In panel (a) the finite-size correlation length $\xi(L,T)/L$ as a
function of the temperature $T$ for different system sizes $L$ is
shown. The data for $L\geq6$ cross at $T_c = 1.042(5)$.  In panel
(b) the corresponding data for the Binder cumulant $g(L,T)$ as a
function of the temperature $T$ for different system sizes $L$ are
shown. The data cross cleanly at $T_c = 1.0385(25)$. The crossing
of the data for the Binder cumulant is again more precise than the
one of the correlation length and shows no noticeable drift of the
crossing point with increasing system sizes.}
\label{fig:Tc_pmJdil}
\end{figure*}

\begin{figure*}
\includegraphics[width=0.95\columnwidth]{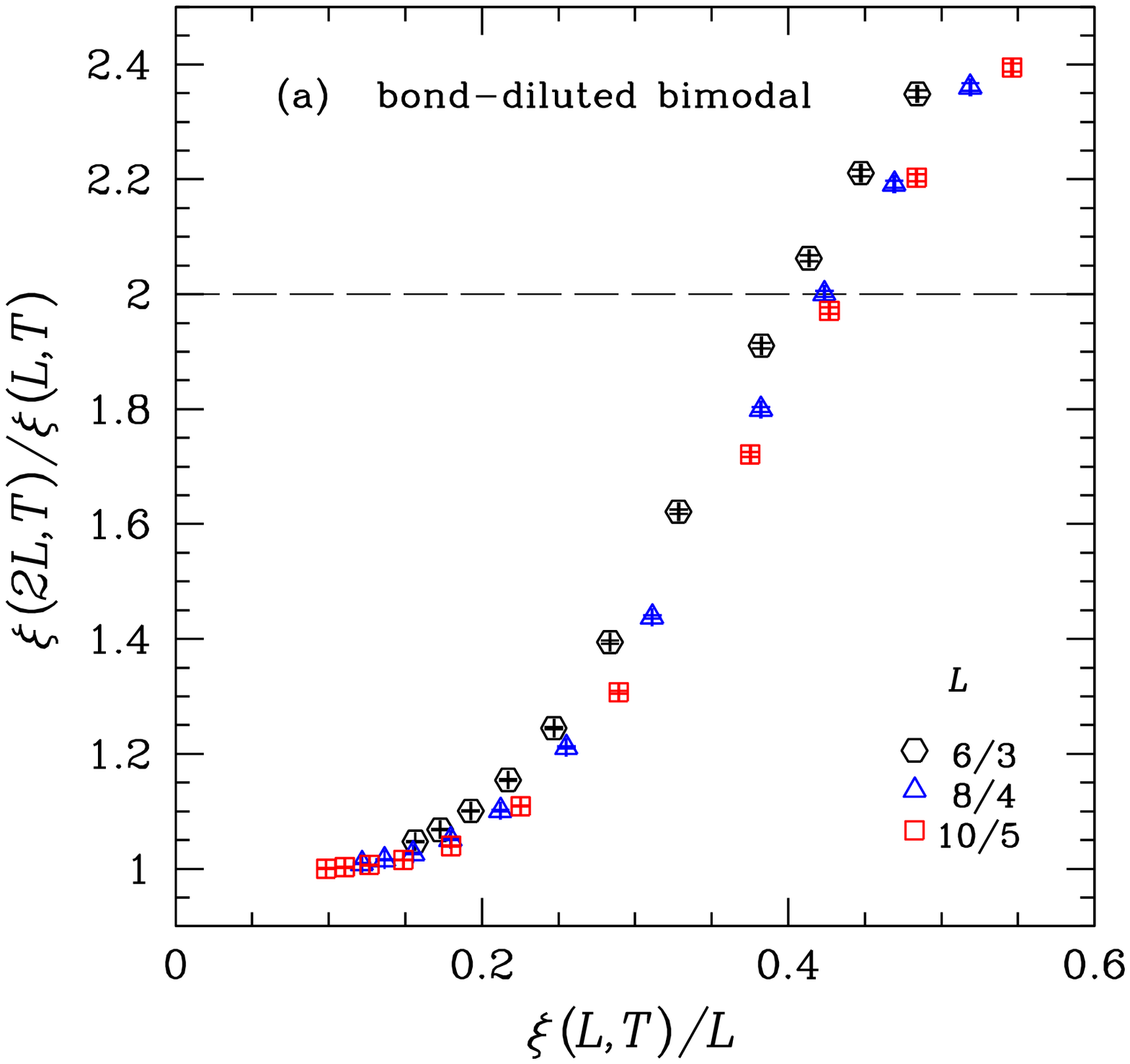}\quad
\includegraphics[width=0.95\columnwidth]{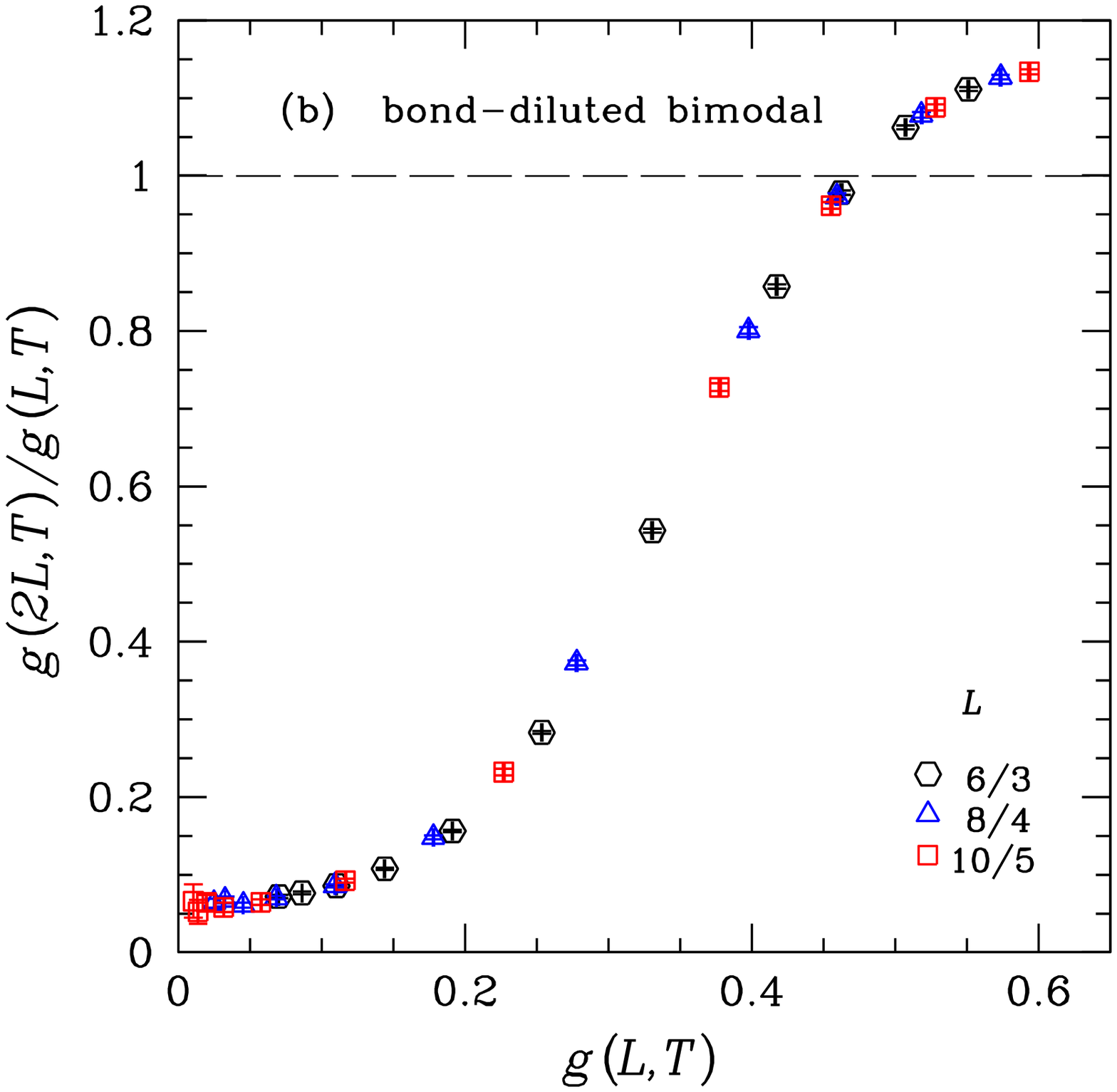}
\vspace*{-1.2cm}
\caption{(Color online)
Finite-size scaling functions of the correlation length $F_\xi$ and
the Binder cumulant $F_g$ of the four-dimensional Edwards-Anderson
Ising spin glass with bond-diluted bimodal disorder for different
system sizes $L$. The comparison between the scaling function $F_\xi$
as a function of the finite-size correlation length $\xi(L,T)/L$
is shown in panel (a) and $F_g$ as a function of the Binder cumulant
$g(L,T)$ shown in panel (b). Given the small system sizes studied,
the Binder cumulant is more suited as a scaling variable, since it
displays clearly smaller scaling corrections than the finite-size
correlation length. The data collapse for the scaling function $F_g$
is excellent for the small system sizes studied. The broken lines in
both panels indicate the condition that defines the effective critical
point $T_c^*$.}
\label{fig:FSS_ratio_pmJdil}
\end{figure*}

\subsection{Gaussian disorder (no dilution)}
\label{subsec:gauss}

Panel (a) of Fig.~\ref{fig:Tc_gauss} shows the data for the finite-size
correlation length as a function of the temperature for different
system sizes $L$. With increasing $L$, the data show a shift of
the effective $T_c$ toward a smaller value of $T_c$. This effect
has already been observed in studies of the three-dimensional EA
model,\cite{katzgraber:06,joerg:06,palassini:03c} however, in contrast
to the three-dimensional case the range of available lattice sizes in
four dimensions is more restricted and therefore the effect of this
shift is clearly a restriction for a precise determination of $T_c$
using the crossings of $\xi(L,T)/L$. Taking the data for $L\geq6$
and neglecting the remaining scaling corrections, we obtain $T_c =
1.810(15)$.

Panel (b) of Fig.~\ref{fig:Tc_gauss} shows the data for the Binder
cumulant $g(L,T)$ as a function of the temperature for different
system sizes. The data cross cleanly at the critical temperature
of $T_c = 1.805(10)$ and there is no shift as for the crossings of
the finite-size correlation length [panel (a)]. In contrast to the
situation in three dimensions the data splay out very well below $T_c$
making a precise determination of $T_c$ from the Binder cumulant
data possible, although the error bars on the Binder cumulant
are slightly larger than the ones on the correlation length. We
consider Fig.~\ref{fig:Tc_gauss} as a first indication that it might
be profitable to use $g(L,T)$ instead of $\xi(L,T)/L$ as scaling
variable in the finite-size scaling analysis in four dimensions,
possibly because $T_c$ is considerably larger and thus the crossing
point further away from $T = 0$ where $g\to1$.

In Table \ref{tab:res_quotient} (Appendix \ref{app:quotient}) we present
the results for the critical quantities we obtain from the quotient
method using $T_c^*$ defined from the crossings of $\xi/L$ and $g$,
respectively. While the results for $T_c$, $g(T_c)$, and $\nu$ show no
noticeable scaling corrections, the ones for $\eta$, $\xi(L,T_c)/L$,
and in a minor extent $\gamma$ indicate clearly the presence of such
corrections. In conclusion we obtain the following values for the
critical quantities for the undiluted Gaussian disorder:
\begin{eqnarray}
\label{eq:results_gauss}
&&T_c = 1.805(10),	\nonumber \\ 
&&g(T_c) = 0.470(5),  	\nonumber \\
&&\nu = 1.02(2),	\nonumber \\
&&\eta = -0.275(25),	\nonumber \\
&&\gamma = 2.32(8). 
\end{eqnarray}

Our value for $\eta$ contains a crude extrapolation to the
thermodynamic limit, which makes use of the fact that corrections to
scaling seem to be disappearing very fast with increasing system
size. The result we give for $\eta$ is justified later in
Fig.~\ref{fig:Fchi_convergence}. The value for $\gamma$ is determined
from our estimates of $\nu$ and $\eta$. Its value depends much more on
a precise determination of $\nu$ than on a precise determination of
$\eta$. We did not try to determine an infinite-volume extrapolation
for the value of $\xi(L,T_c)/L$, which is another universal quantity,
because the data from such a limited range of system sizes do not
allow for a controlled extrapolation.  Note that in the case of $\eta$,
where one might expect similar problems, we rely on the convergence of
the estimates from opposing sides for the different scaling functions
(see Fig.~\ref{fig:Fchi_convergence}) thus allowing for a somewhat more
reliable extrapolation. Taking into account the presence of corrections
to scaling, the estimates for the critical exponents obtained from the
extended scaling method of Campbell {\em et al.}\cite{campbell:06}
(see Appendix \ref{app:campbell}) do agree with our final estimates.

\begin{figure}
\includegraphics[width=0.95\columnwidth]{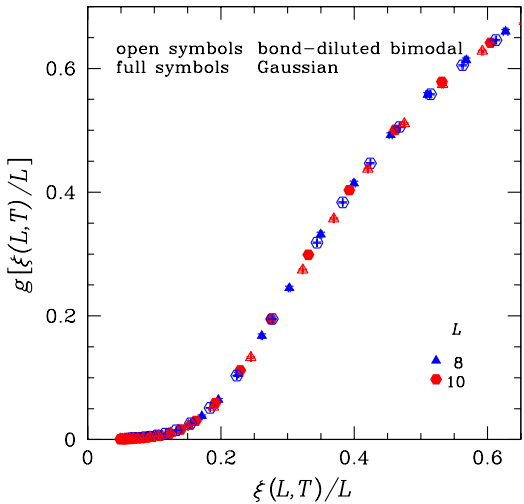}
\vspace*{-1.0cm}
\caption{(Color online) 
Comparison of the Binder cumulant $g$ as a function of the finite-size
correlation length for the four-dimensional EA Ising spin glass with
Gaussian (full symbols) and diluted bimodal disorder (open symbols)
for different system sizes $L \geq 8$. Both functions agree very well; 
further evidence for universal behavior.}
\label{fig:FSS_g_vs_xi_4D}
\end{figure}

\begin{figure*}[htbf]
\includegraphics[width=0.95\columnwidth]{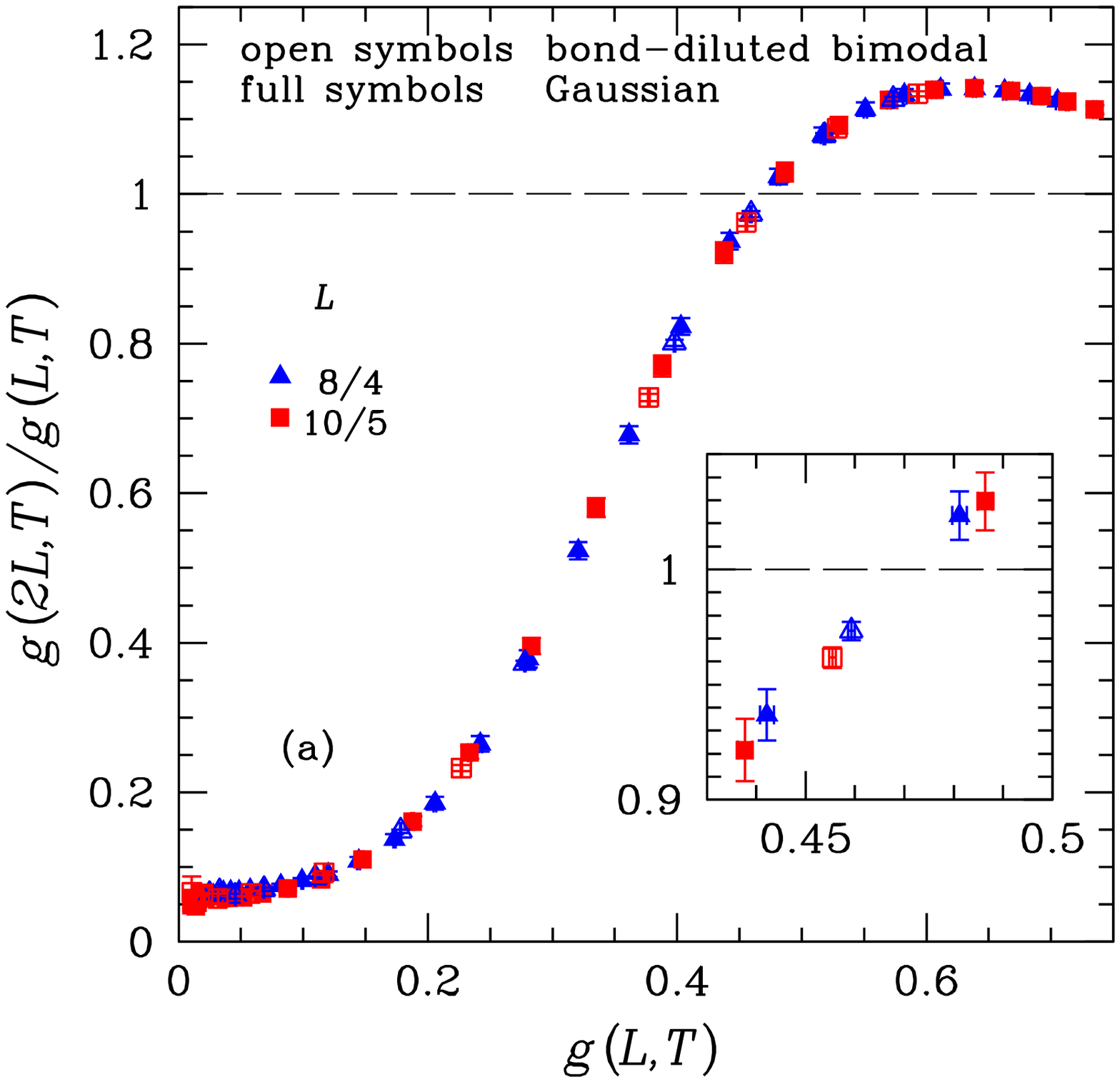}\quad
\includegraphics[width=0.95\columnwidth]{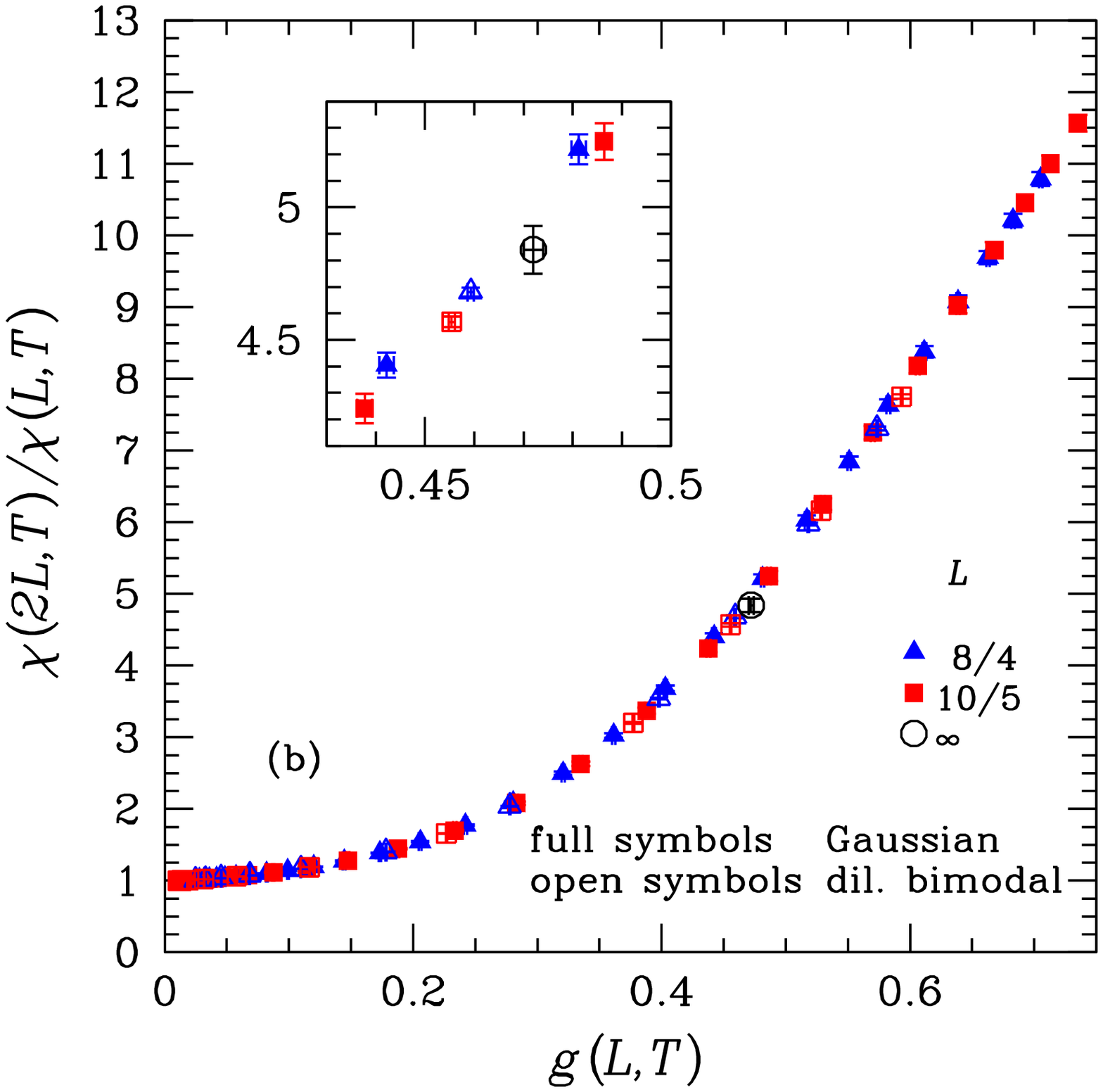}
\vspace*{-1.2cm}
\caption{(Color online) 
Comparison of the finite-size scaling function $F_g$ shown in panel
(a) and $F_\chi$ shown in panel (b) for the four-dimensional model
with Gaussian and bond-diluted bimodal disorder for different system
sizes $L$ as a function of the Binder cumulant $g$. The data for both
finite-size scaling functions show very little scaling corrections.
The fact that the curves for the two different models for $F_g$ and
$F_\chi$ fall on one single curve is a strong evidence for universal
critical behavior of the four-dimensional Edwards-Anderson model. The
insets in both panels present an enlarged view around the critical
point [which is at $g(L,T_c)=0.472(2)$] and show in more detail
that the data for the finite-size scaling functions collapse onto one
single master curve. The data point with the label ``$\infty$'' in panel
(b) indicates our infinite-volume extrapolation of $F_\chi$ at
criticality corresponding to $\eta=0.275(25)$ and $g(T_c)=0.472(2)$.}
\label{fig:FSS_ratio_comparison}
\end{figure*}

\subsection{Diluted bimodal disorder}
\label{subsec:pmj}

Panel (a) in Fig.~\ref{fig:Tc_pmJdil} shows data for the finite-size
correlation length as a function of the temperature for different
system sizes. With increasing system size $L$ the data show---as
for the Gaussian data---a noticeable shift of the effective
critical temperature $T_c$ toward a smaller value.  Taking the
data for $L\geq6$ and neglecting the remaining scaling corrections
we obtain $T_c = 1.042(5)$.  Panel (b) in Fig.~\ref{fig:Tc_pmJdil}
shows the data for the Binder cumulant $g(L,T)$ as a function of the
temperature for different system sizes. The data cross cleanly, and
we determine $T_c=1.0385(25)$. We find again as in the Gaussian case
that the Binder cumulant has smaller corrections to scaling than the
correlation length.  This is confirmed by a comparative analysis of the
finite-size scaling functions $F_{\xi}$ as a function of $\xi(L,T)/L$
and $F_g$ as a function of $g(L,T)$.

The finite-size scaling functions $F_\xi$ and $F_g$ contain in
principle the very same information. However, due to the fact that we
are working in a restricted range of lattice sizes there are visible
differences.  In panel (a) of Fig.~\ref{fig:FSS_ratio_pmJdil}
we show the scaling function $F_\xi$ and in panel (b) of
Fig.~\ref{fig:FSS_ratio_pmJdil} of the corresponding scaling function
$F_g$. Comparing the two figures one clearly sees that $F_g$ shows
much smaller finite-size corrections than $F_\xi$. This figure is clear
evidence that for our data the Binder cumulant $g(L,T)$ is better suited
as a scaling variable than the correlation length $\xi(L,T)/L$. This
is in contrast to recent work in three dimensions, where typically
$\xi(L,T)/L$ is used instead of $g(L,T)$ in order to cleanly
determine $T_c$ and the critical exponents.\cite{ballesteros:00,
katzgraber:06} The reason for this behavior is possibly given by the
following arguments: First, in three dimensions the crossing of the
Binder cumulant data is at a rather flat angle making it difficult to
determine $T_c$ precisely, while in four dimensions the crossing is at
a much steeper angle. Second, the range of available system sizes in
four dimensions is smaller than in three dimensions and therefore the
use of observables with small scaling violations is better in order
to have good control over the estimates of the critical quantities.

In Table \ref{tab:res_quotient} (Appendix \ref{app:quotient}) we
present the results for the critical quantities we have obtained from the
quotient method using $T_c^*$, defined either from the crossings of
$\xi/L$ or $g$, respectively. We obtain the following values for the
critical quantities for the link-diluted bimodal disorder distribution:
\begin{eqnarray}
\label{eq:results_pmJdil}
&&T_c = 1.0385(25),	\nonumber \\
&&g(T_c) = 0.472(2), 	\nonumber \\
&&\nu = 1.025(15), 	\nonumber \\
&&\eta = -0.275(25),	\nonumber \\\
&&\gamma = 2.33(6).
\end{eqnarray}
The estimate of $\eta$ is justified below in
Fig.~\ref{fig:Fchi_convergence}.

\subsection{Comparison of finite-size scaling functions}
\label{subsec:comparison}

\begin{figure}
\includegraphics[width=0.95\columnwidth]{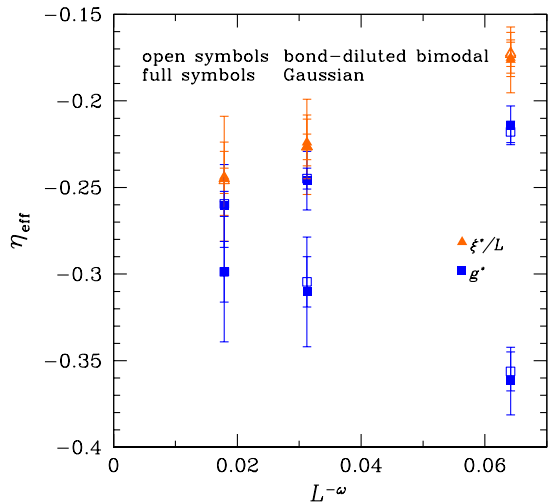}
\vspace*{-1.0cm}
\caption{(Color online) 
Convergence of the effective exponent $\eta_{\rm eff}$ defined
from $F_\chi$ and $F_\zeta$ through Eqs.~(\ref{eq:Fchi}) and
(\ref{eq:Fzeta}), respectively, at $T_c^*$ (defined by $\xi^*/L$ and
$g^*$) as a function of $L^{-\omega}$ with $\omega=2.5$. The data for
the Gaussian (full symbols) and bond-diluted bimodal (open symbols)
disorder are consistent within errors and they are also consistent
with a unique value of $\eta = -0.275(25)$ for an infinite-volume
extrapolation.}
\label{fig:Fchi_convergence}
\end{figure}

\begin{figure*}
\includegraphics[width=0.95\columnwidth]{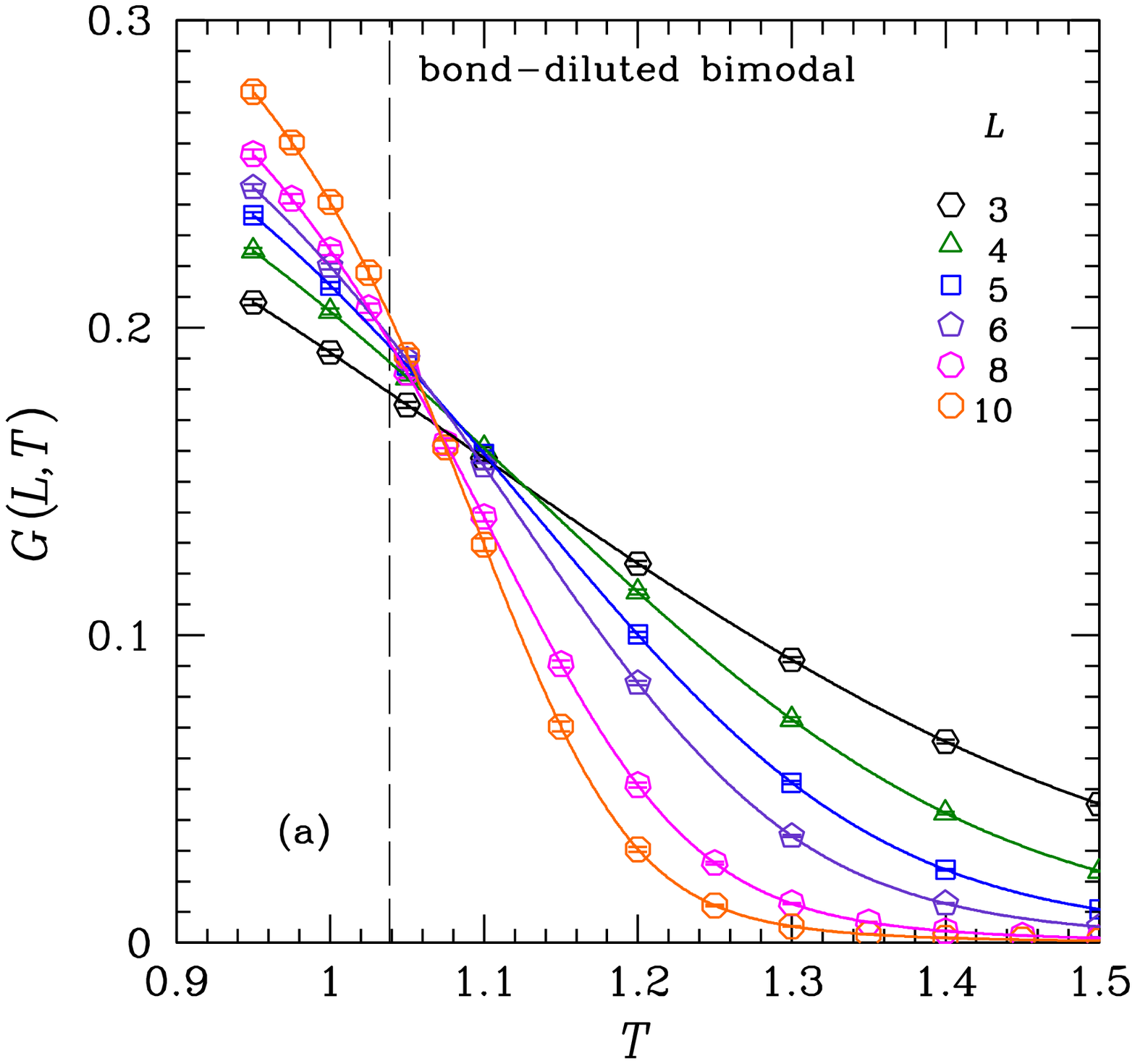}\quad
\includegraphics[width=0.95\columnwidth]{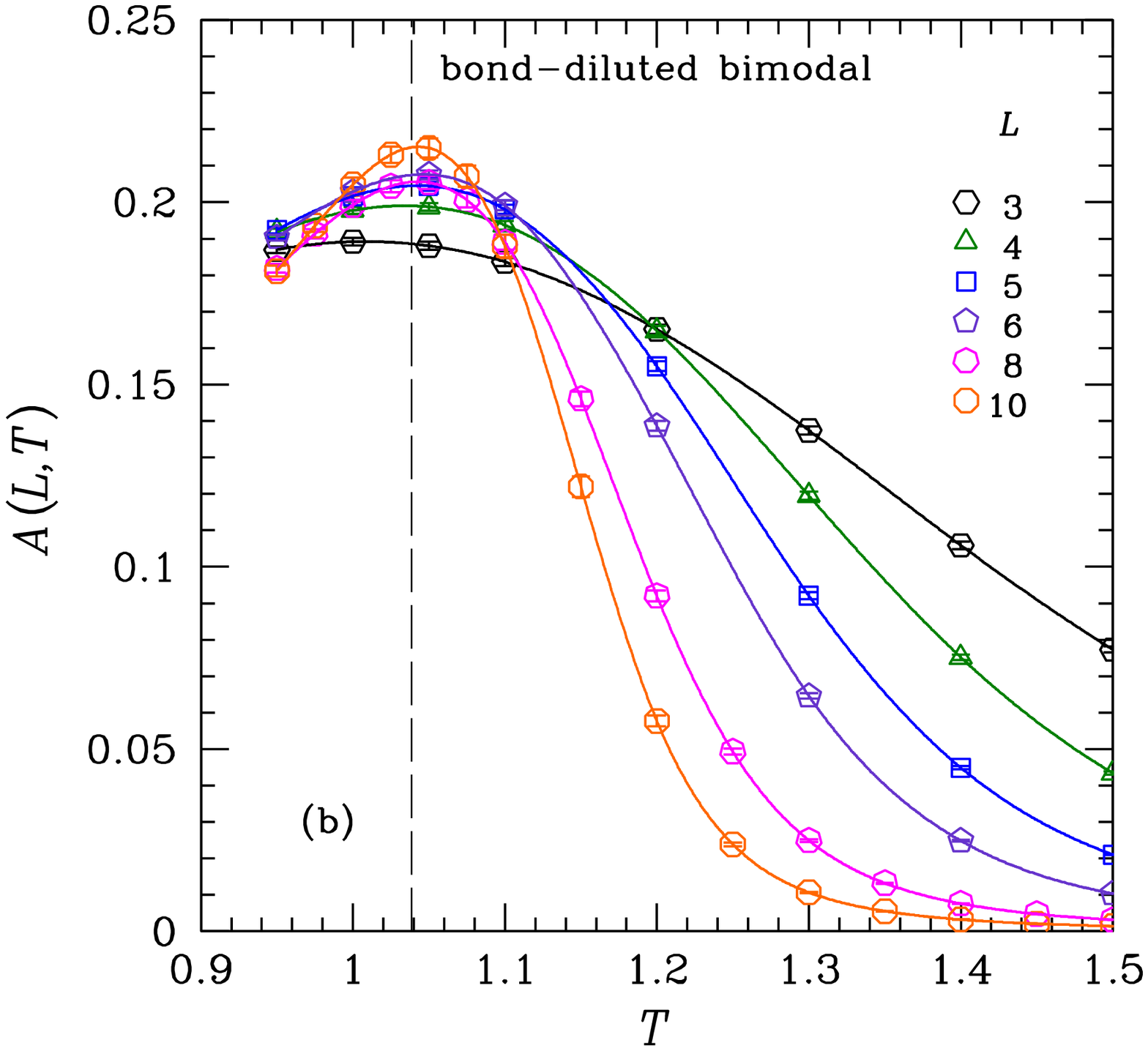}
\vspace*{-1.2cm}
\caption{(Color online)
The Guerra parameter $G(L,T)$ [panel (a)] and the lack of
self-averaging parameter $A(L,T)$ [panel (b)] as a function of
temperature $T$ for the four-dimensional Edwards-Anderson model
with bond-diluted bimodal disorder for different system sizes $L$.
The crossings shift noticeably to a smaller effective $T_c$ in both
cases for increasing system size making an accurate determination of
$T_c$ from these data difficult and imprecise. In the case of $A(L,T)$
the fact that the crossing occurs close to the maximum of the curves
makes the situation for an accurate determination of $T_c$ (vertical
dashed lines) impossible.}
\label{fig:G_and_A_pmJdil}
\end{figure*}

\begin{figure*}
\includegraphics[width=0.95\columnwidth]{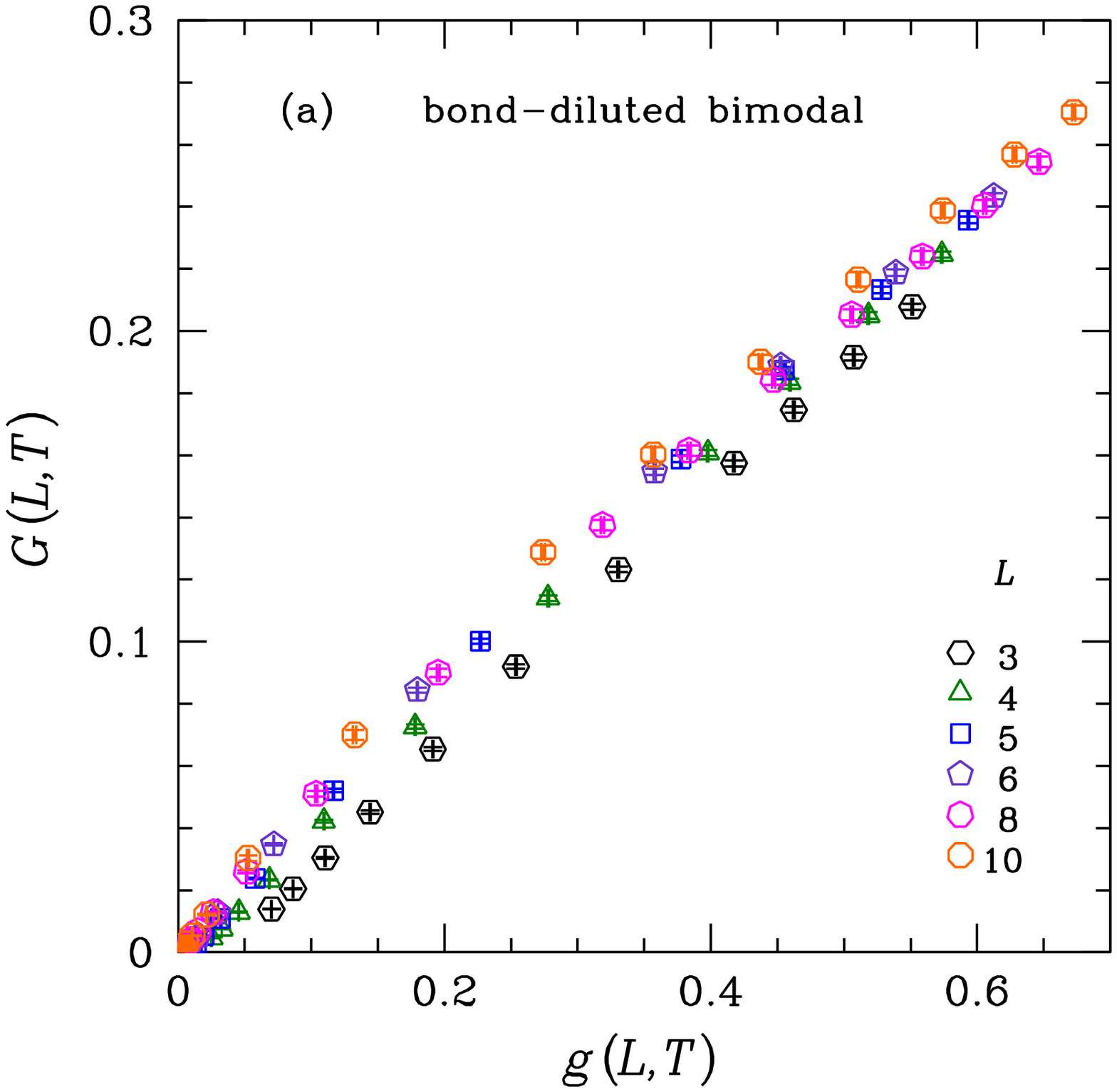}\quad
\includegraphics[width=0.95\columnwidth]{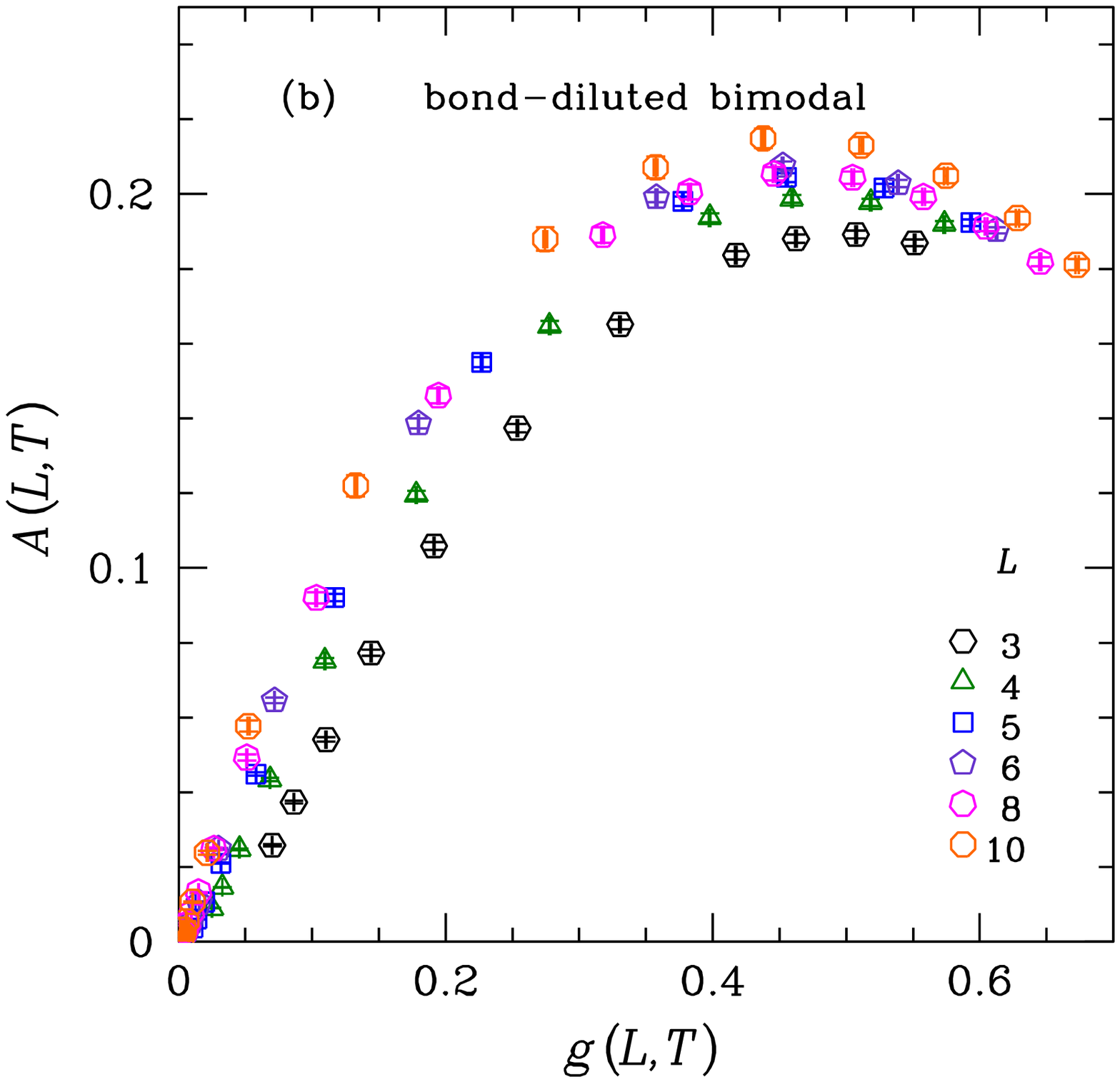}
\vspace*{-1.2cm}
\caption{(Color online)
The Guerra parameter $G(L,T)$ [panel (a)] and the lack of
self-averaging parameter $A(L,T)$ [panel (b)] as a function of the
Binder cumulant $g(L,T)$ for the four-dimensional Edwards-Anderson
Ising spin glass with bond-diluted bimodal disorder for different
system sizes $L$. Both $G(L,T)$ and $A(L,T)$ display large corrections
to scaling.}
\label{fig:G_and_A_vs_g_pmJdil}
\end{figure*}

The results for the critical quantities given in
Eqs.~(\ref{eq:results_gauss}) and (\ref{eq:results_pmJdil}) are
consistent with universal critical behavior of the four-dimensional
Edwards-Anderson Ising spin-glass model.  We further strengthen this
result by comparing the finite-size scaling functions of the two
models not only at the critical point but within the whole scaling
region. The direct comparison of the finite-size scaling functions
has shown to be probably the best approach to check for universality
in spin glasses as it allows for a completely parameter-free
comparison of different models (see Refs.~\onlinecite{joerg:06} and
\onlinecite{katzgraber:06} for a comparison of different models
in three space dimensions). In Fig.~\ref{fig:FSS_g_vs_xi_4D}
we compare the Binder cumulant $g(L,T)$ plotted against the
correlation length $\xi(L,T)/L$ for the two different disorder
distributions. The data agree and collapse onto a single master
curve, which is a strong evidence for universality. In panel (a) of
Fig.~\ref{fig:FSS_ratio_comparison} we compare the finite-size scaling
function $F_g$ as a function of the Binder cumulant $g(L,T)$. The data
collapse again within error bars onto a single master curve. This
result illustrates once more that the estimates of the critical
exponent $\nu$ of the two disorder distributions do agree within
error bars. In panel (b) of Fig.~\ref{fig:FSS_ratio_comparison}
we compare the finite-size scaling function $F_\chi$ as a function
of the Binder cumulant $g(L,T)$. Again the data collapse is within
the error bars, which means that also the estimates of the critical
exponent $\eta$ of the two models coincide within errors.

Finally, in Fig.~\ref{fig:Fchi_convergence} we show the convergence of
the effective exponent $\eta_{\rm eff}$ from $F_\chi$ and $F_\zeta$
through Eqs.~(\ref{eq:Fchi}) and (\ref{eq:Fzeta}), respectively,
measured at the crossings of $\xi/L$ and $g$ as a function
of $L^{-\omega}$. All the scaling corrections are compatible
with a leading correction-to-scaling exponent of $\omega \approx
2.5$. Clearly, this is only an effective exponent because for the small
system sizes we have at hand, we cannot expect to be in the asymptotic
scaling regime where the leading correction-to-scaling exponent
dominates. However, our estimate of $\omega$ is consistent with the
assumption that the corrections to scaling we see are dominated by
the first nonleading correction-to-scaling term since our value
for $\omega$ is roughly twice as large as the one obtained from
high-temperature expansion studies in Ref.~\onlinecite{daboul:04} or
alternatively they might be due to remaining analytic corrections with 
an effective correction-to-scaling exponent given by $2-\eta$.
Figure \ref{fig:Fchi_convergence} shows that an infinite-volume
extrapolation of the different estimates is compatible with a unique
value of $\eta = -0.275(25)$ in the infinite-volume limit for {\em
both} models.

\section{Conclusions}
\label{sec:conclusions}

We have tested universality in four-dimensional Ising spin glasses and
computed precise estimates of the critical parameters of the model with
both Gaussian and link-diluted bimodal disorder.  Our results show that
the different critical exponents for different disorder distributions
agree well. Furthermore, by plotting the Binder ratio as a function
of the correlation length, we show that four-dimensional Ising spin
glasses (with compact disorder distributions) seem to share the same
universality class.  Furthermore, we compute different finite-size
scaling functions in four space dimensions defined via ratios of
different observables and show that these show small corrections to
scaling, especially when studied as a function of the Binder parameter.
The results presented thus indicate that universality is not violated
in four-dimensional spin glasses.

\begin{acknowledgments}

We would like to thank I.~A.~Campbell and A.~P.~Young for the
helpful discussions.  The simulations have been performed in part on
the Brutus, Hreidar, and Gonzales clusters at ETH Z\"urich and on
the Piovra cluster at the Universit\`{a} di Roma ``La Sapienza.''
We would like to thank in particular O.~Byrde for providing
beta-testing access to the Brutus cluster.  T.J.~acknowleges the
support from the EEC's HPP HPRN-CT-2002-00307 (DYGLAGEMEM) and the FP6
IST program under Contract No.~IST-034952 (GENNETEC). The LPTMS is an
Unit\'e de Recherche de l'Universit\'e Paris~XI associ\'ee au CNRS.
H.G.K.~acknowledges the support from the Swiss National Science
Foundation under Grant No.~PP002-114713.

\end{acknowledgments}

\begin{table*}
\caption{
Results from the quotient method for the Gaussian (top) and
link-diluted bimodal (bottom) disorder at $T_c^*$ determined from
the crossings of $\xi/L$ and $g$, respectively.  We group the data
according to the observables which were used to compute them (separated
by vertical bars).
\label{tab:res_quotient}
}
\begin{tabular*}{\textwidth}{@{\extracolsep{\fill}} l c | c | c c c| c c c | c
c c}
  \hline
  \hline
  Crossings of $\xi/L$ & $L$ & $T_c^*$ & $\xi^*/L$ & $\partial_\xi F_\xi^*$
  & $\nu_{\rm eff}$ & $F_\chi^*$ & $\eta_{\rm eff}$  & $\gamma_{\rm eff}$ &
$F_\zeta^*$ &
  $\eta_{\rm eff}$ & $\gamma_{\rm eff}$ \\
  \hline
  &$3/6$  & $1.858(6)$ & $0.396(2)$ & $5.00(3)$ & $1.007(8)\phantom{1}$ &
  $4.52(6)\phantom{1}$ & $-0.18(2)$ & $2.20(3)\phantom{1}$ &
$1.129(7)\phantom{1}$ & $-0.175(9)\phantom{1}$ & $2.19(3)\phantom{1}$ \\
  &$4/8$  & $1.824(7)$ & $0.418(3)$ & $4.67(5)$ & $1.018(14)$ &
  $4.68(9)\phantom{1}$ & $-0.23(3)$ & $2.27(6)\phantom{1}$ & $1.168(11)$ &
$-0.224(16)$ & $2.26(5)\phantom{1}$ \\
  &$5/10$ & $1.814(8)$ & $0.427(4)$ & $4.62(6)$ & $1.010(17)$ &
  $4.74(12)$ & $-0.24(4)$ & $2.26(8)\phantom{1}$ & $1.184(12)$ & $-0.244(15)$
& $2.27(5)\phantom{1}$ \\ 
  \hline 
  Crossings of $g$ & $L$ & $T_c^*$ & $g^*$  & $\partial_g F_g^*$  &
  $\nu_{\rm eff}$ & $F_\chi^*$ & $\eta_{\rm eff}$ & $\gamma_{\rm eff}$ &
$F_\zeta^*$ &
  $\eta_{\rm eff}$ & $\gamma_{\rm eff}$ \\ 
  \hline 
  &$3/6$  & $1.803(6)$ & $0.470(3)$ & $2.01(5)\phantom{1}$ & $1.042(25)$ &
  $5.14(7)\phantom{1}$ & $-0.36(2)$ & $2.46(8)\phantom{1}$ &
$1.160(9)\phantom{1}$ & $-0.214(12)$ & $2.31(7)\phantom{1}$ \\
  &$4/8$  & $1.805(8)$ & $0.469(5)$ & $2.08(9)\phantom{1}$ & $1.018(42)$ &
  $4.96(11)$ & $-0.31(3)$ & $2.35(13)$ & $1.186(14)$ & $-0.246(17)$ &
$2.29(11)$ \\
  &$5/10$ & $1.805(8)$ & $0.471(7)$ & $2.06(11)$           & $1.022(54)$ &
  $4.92(14)$ & $-0.30(4)$ & $2.35(17)$ & $1.198(20)$ & $-0.261(24) $ &
$2.31(15)$ \\
  \hline
  \hline \\
  \hline
  \hline
  Crossings of $\xi/L$ & $L$  & $T_c^*$ & $\xi^*/L$ & $\partial_\xi F_\xi^*$ &
  $\nu_{\rm eff}$ & $F_\chi^*$ &  $\eta_{\rm eff}$ & $\gamma_{\rm eff}$ &
$F_\zeta^*$ & $\eta_{\rm eff}$ & $\gamma_{\rm eff}$\\
  \hline
  & $3/6$  & $1.0716(15)$ & $0.400(2)$ & $4.86(2)$ & $1.021(7)\phantom{1}$
  & $4.51(4)$ & $-0.17(1)$ & $2.22(3)$ & $1.127(6)$ & $-0.172(8)$ & $2.22(3)$
\\
  & $4/8$  & $1.0502(16)$ & $0.424(2)$ & $4.56(2)$ & $1.025(7)\phantom{1}$
  & $4.68(6)$ & $-0.23(2)$ & $2.29(3)$ & $1.170(6)$ & $-0.227(8)$ & $2.28(3)$
\\
  & $5/10$ & $1.0441(18)$ & $0.433(2)$ & $4.42(3)$ & $1.032(9)\phantom{1}$
  & $4.74(7)$ & $-0.25(2)$ & $2.32(4)$ & $1.186(6)$ & $-0.246(8)$ & $2.32(3)$
\\
  \hline
  Crossings of $g$ & $L$ & $T_c^*$& $g^*$ & $\partial_g F_g^*$ & $\nu_{\rm
    eff}$ & $F_\chi^*$ & $\eta_{\rm eff}$ & $\gamma_{\rm eff}$ & $F_\zeta^*$ &
$\eta_{\rm eff}$ & $\gamma_{\rm eff}$ \\
  \hline
  & $3/6$  & $1.0390(18)$ & $0.472(2)$ & $2.09(3)$ & $1.010(14)$ & $5.12(4)$
  & $-0.36(1)$ & $2.38(3)$ & $1.163(5)$ & $-0.224(7)$ & $2.25(3)$ \\
  & $4/8$  & $1.0390(18)$ & $0.472(2)$ & $2.06(3)$ & $1.021(15)$ & $4.94(5)$
  & $-0.31(2)$ & $2.36(5)$ & $1.185(5)$ & $-0.245(7)$ & $2.29(4)$ \\
  & $5/10$ & $1.0384(20)$ & $0.473(3)$ & $2.06(4)$ & $1.019(20)$ & $4.92(6)$
  & $-0.30(2)$ & $2.34(7)$ & $1.197(6)$ & $-0.259(8)$ & $2.30(6)$ \\
  \hline
  \hline
\end{tabular*}
\end{table*}

\appendix

\begin{figure}
\includegraphics[width=0.95\columnwidth]{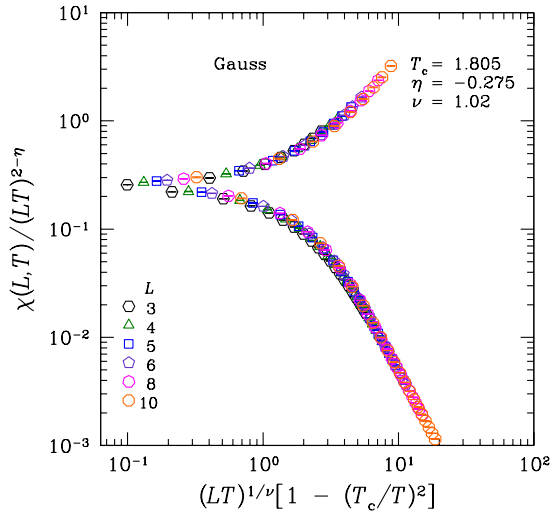}
\vspace*{-1.2cm}
\caption{(Color online)
Scaling plot of the spin-glass susceptibility according to
the extended scaling approach for the system with Gaussian
disorder. We use the estimates of the critical exponents presented
in Eqs.~(\ref{eq:results_gauss}). The data for the susceptibility
scale very well with our estimates of the critical parameters.}
\label{fig:chi_extended_gauss_4D}
\end{figure}

\section{Other observables}
\label{sec:appendix}

In this appendix we discuss the quantities that have been shown
to be less useful in the study of the location of the spin-glass
transition and the issue of the universality of the four-dimensional
Edwards-Anderson model. We have decided to present our results
concerning these quantities since our data for the bond-diluted bimodal
coupling distribution has by far the best statistics in the context of
Monte Carlo simulations of the four-dimensional Edwards-Anderson model.

Panel (a) of Fig.~\ref{fig:G_and_A_pmJdil} shows the data
for the Guerra parameter $G(L,T)$ defined in Eq.~(\ref{eq:Guerra})
as a function of the temperature for different system sizes. The
crossings of the data for increasing system sizes shift noticeably
toward smaller temperatures.  We find that $G(L,T)$ has rather large
finite-size corrections and also relatively large errors compared to,
e.g., the Binder cumulant and therefore is not well suited for an
accurate determination of $T_c$.  The same conclusions have been found
by Ballesteros {\em et al}.~\cite{ballesteros:00} for the three-dimensional
bimodal Ising spin glass. Note, however, that in the case of mean-field
spin glasses the situation may be different as the Guerra parameter
has shown to be more efficient in locating a spin-glass transition
than the Binder cumulant in certain situations.\cite{palassini:03c}

Panel (b) of Fig.~\ref{fig:G_and_A_pmJdil} shows the data for the
lack of self-averaging parameter $A(L,T)$ defined in Eq.~(\ref{eq:A})
as a function of the temperature for different system sizes. This
parameter is related to the Guerra parameter and also shows large
finite-size corrections and due to the fact that the crossings happen
close to the maximum of the curves, where the slope changes very
fast, the crossings (apart from the fact that they move noticeably)
cannot be determined reliably. This fact together with the
rather large relative error makes that this quantity the least
suited for a precise determination of the critical temperature of
all the quantities discussed here.

In Fig.~\ref{fig:G_and_A_vs_g_pmJdil} we show the data for $G(L,T)$
[panel (a)] and $A(L,T)$ [panel (b)] as a function of the Binder
cumulant. Clearly, the two quantities have strong finite-size
corrections.  The fact that these quantities may present such strong
finite-size scaling corrections should be kept in mind when, e.g.,
the behavior of $A(L,T)$ is used to make statements on the nature of
the spin-glass phase as it is done in Ref.~\onlinecite{palassini:03c}.

\section{Derivation of the quotient relations}
\label{app:derivation}

In the following, we derive Eqs.~\eqref{eq:quotient_2} and
\eqref{eq:quotient_3} in detail.  Starting from the definition of
$F_{\xi}$, we can write,
\begin{equation}
  \label{eq:der1}
   \xi(sL, T) = F_{\xi}[\xi(L,T)/L;s] \xi(L,T).
\end{equation}
Taking the derivative with respect to $T$, using the chain rule and finally
dividing by $\partial_T \xi(L,T)$, we arrive at 
\begin{equation}
  \label{eq:der2}
  \frac{\partial_T \xi(sL,T)}{\partial_T \xi(L,T)} = \frac{\xi(L,T)}{L} \partial_x
  F_{\xi}(x;s) + F_{\xi}[\xi(L,T)/L]. 
\end{equation}
Note that 
\begin{equation}
  \label{eq:der3}
  F_{\partial_T \xi}[\xi(L,T)/L;s] = \frac{\partial_T \xi(sL,T)}{\partial_T
    \xi(L,T)}.
\end{equation} 
Using the fact that at the effective critical point, where
$\xi(L,T_c^*)/L=x^*$, we have $F_{\xi}(x^*;s)=s$ and the fact that
the correlation length $\xi$ close to the critical point has a simple
scaling form given in Eq.~(\ref{eq:xi_scale}), we see that
\begin{align}
  \label{eq:der4}
  F_{\partial_T \xi}(\xi(L,T)/L;s) &= s^{1/\nu + 1} + O(L^\omega) \notag \\
  &= x^* \partial_x F_{\xi}(x;s) \big|_{x=x^*} + s,
\end{align}
from which we obtain Eq.~\eqref{eq:quotient_2} by dividing by $s$.
The derivation of Eq.~\eqref{eq:quotient_3} has as the starting point
\begin{equation}
  \label{eq:der5}
   g(sL, T) = F_{g}(g(L,T);s) g(L,T),
\end{equation}
and for the rest is analogous to the one given for
Eq.~\eqref{eq:quotient_2}.

\section{Results from the quotient method}
\label{app:quotient}

In this section we list the detailed results from the quotient method.
The data are grouped by the observable used to compute the estimates
in Table \ref{tab:res_quotient}.

\section{Extended scaling}
\label{app:campbell}

Recently, Campbell {\em et al.}\cite{campbell:06} suggested an extended
scaling approach, which allows one to extend the scaling region from
$|L^{1/\nu}(T - T_c)| \lesssim 1$ to virtually $T \to \infty$.  The
method has the advantage in that it drastically reduces the corrections
to scaling commonly found when performing a simple finite-size scaling
analysis of the spin-glass susceptibility.\cite{katzgraber:06} In
that scaling approach the scaling equation for the susceptibility
[see Eq.~(\ref{eq:chisg_scale})] is modified in the following
way:\cite{comment:campbell1}
\begin{equation}
\chi_{\rm SG}(L,T) \sim (LT)^{2 - \eta}\
			\tilde{\chi}[B(LT)^{1/\nu}|1 - (T_c/T)^2|] .
\label{eq:extended_chi}
\end{equation}

In Fig.~\ref{fig:chi_extended_gauss_4D} we illustrate the quality of
the critical parameters by performing an extended finite-size scaling
plot of the susceptibility for the EA spin glass with Gaussian disorder
obtained from the quotient method.\cite{ballesteros:97,ballesteros:00}
Similar results are obtained for the model with bond-diluted bimodal
disorder, as well as other observables.

\bibliography{refs,comments}

\begin{thebibliography}{62}
\bibitem[{com({\natexlab{a}})}]{comment:dilute_FP}
\bibinfo{note}{The presence of such an additional fixed-point at zero
  temperature is also found in the case of diluted spin-glass models with
  discrete coupling distributions as discussed in Ref.~\onlinecite{joerg:08}.}
\bibitem[{com({\natexlab{b}})}]{comment:wang}
\bibinfo{note}{Note that Wang and Swendsen also estimated $\nu \approx 0.66(7)$
  using a Monte Carlo renormalization group method. See
  Ref.~\onlinecite{wang:88a} for details.}
\bibitem[{com({\natexlab{c}})}]{comment:dil}
\bibinfo{note}{Note that here we have adopted the common practice in diluted
  systems which is that the variance of the distribution is {\it not}
  normalized to unity.}
\bibitem[{com({\natexlab{d}})}]{comment:campbell2}
\bibinfo{note}{Another possibility is to use the extended scaling approach
  presented in Ref.~\onlinecite{campbell:06} where analytic corrections to
  scaling are also avoided. We have analyzed our data with this method
  obtaining results which agree very well with the quotient method
  (Ref.~\onlinecite{ballesteros:97}).}
\bibitem[{com({\natexlab{e}})}]{comment:campbell1}
\bibinfo{note}{Similar changes apply for other observables such as the
  correlation length or the Binder ratio. Since both are dimensionless, the
  finite-size scaling equation for these observables ${\mathcal O}$ changes to
  ${\mathcal O} = \tilde{\mathcal O}[(LT)^{1/\nu}|1 - (T_c/T)^2|]$.}
\end{thebibliography}

\end{document}